\documentclass[aps,prl,twocolumn,superscriptaddress,nobibnotes]{revtex4-2}

\usepackage[english]{babel}
\usepackage{newtxtext}
\usepackage{lettrine}
\usepackage{enumitem}

\usepackage{newtxmath}
\usepackage{stmaryrd}     
\usepackage{bbm}          

\usepackage{graphicx}
\usepackage[usenames,dvipsnames]{xcolor}
\usepackage{xspace}

\usepackage{varioref}
\usepackage[colorlinks=false]{hyperref}
\usepackage{cleveref}

\usepackage{siunitx}
\usepackage{algorithm}
\usepackage{algorithmic}
\newcommand{\scRNA}{scRNA\xspace}
\newcommand{\KMC}{kinetic Monte Carlo\xspace}

\newcommand{\dd}{\mathrm{d}}
\newcommand{\ii}{\mathrm{i}}
\newcommand{\ee}{\mathrm{e}}
\newcommand{\ud}[1]{\mathrm{d} #1 \,}
\renewcommand{\vec}[1]{\mathbf{#1}}

\newcommand{\oO}[1]{\mathrm{O}({#1})}

\newcommand{\W}{\mathbbm{W}}
\renewcommand{\P}{\mathbbm{P}}

\newlist{todolist}{itemize}{2}
\setlist[todolist]{label=$\square$}

\newenvironment{figDiff}{\begin{figure}[b]}{\end{figure}}
\newenvironment{figMuellerApplication}{\begin{figure}[t]}{\end{figure}}
\newenvironment{figMuellerTP}{\begin{figure}[h]}{\end{figure}}
\newenvironment{figScrnaApplication}{\begin{figure}[b]}{\end{figure}}
\newenvironment{figScrnaInsights}{\begin{figure*}[h]}{\end{figure*}}

\labelformat{subsection}{\thesection.#1}

\begin{document}
\title{Markov-bridge generation of transition paths and its application to cell-fate choice}

\newcommand{\affilczbiohub}{Chan Zuckerberg Biohub---SF, 499 Illinois Street, San Francisco, CA 94158, USA}

\author{Guillaume Le Treut}
\email[\normalfont To whom correspondence should be addressed:~]{guillaume.letreut@gmail.com}
\affiliation{\affilczbiohub}

\author{Sarah Ancheta}
\affiliation{\affilczbiohub}

\author{Greg Huber}
\email[\normalfont To whom correspondence should be addressed:~]{gerghuber@gmail.com}
\affiliation{\affilczbiohub}

\author{Henri Orland}
\affiliation{Institut de Physique Th\'eorique, CEA, CNRS, Universit\'e Paris-Saclay, 91191 Gif-sur-Yvette, France}

\author{David Yllanes}
\affiliation{\affilczbiohub}
\affiliation{Instituto de Biocomputaci\'on y F\'{\i}sica de Sistemas Complejos (BIFI), 50018 Zaragoza, Spain}

\date{\today}

\begin{abstract}
We present a method to sample Markov-chain trajectories constrained to
both the initial and final conditions, which we term Markov bridges. The
trajectories are conditioned to end in a specific state at a given time.
We derive the master equation for Markov bridges, which exhibits the
original transition rates scaled by a time-dependent factor. Trajectories
can then be generated using a refined version of the Gillespie algorithm.
We illustrate the benefits of our method by sampling trajectories in the
M\"uller-Brown potential. This allows us to generate transition paths
which would otherwise be obtained at a high computational cost with standard
Kinetic Monte Carlo methods because commitment to a transition path is
essentially a rare event.
We then apply our method to a single-cell RNA sequencing dataset from
mouse pancreatic cells to investigate the cell differentiation
pathways of endocrine-cell precursors. By sampling Markov bridges
for a specific differentiation pathway we obtain a
time-resolved dynamics that can reveal features such as cell types which behave as
bottlenecks. The ensemble of trajectories also gives information about the
fluctuations around the most likely path. For example,
we quantify the statistical weights of different branches in the
differentiation pathway to alpha cells.
\end{abstract}



\maketitle

\section{Introduction}
In a world where computational capabilities are continually expanding,
stochastic simulations have become indispensable tools, shedding light on
myriad phenomena across physics, chemistry, and biology. These simulations
are crucial for exploring processes such as the spontaneous folding and
unfolding of proteins~\cite{Shaw2008}, allosteric
transitions~\cite{Elber2011}, glassy dynamics~\cite{Cavagna2009,Charbonneau2023}, the binding of
molecules~\cite{Shan2011}, and more generally for computing physical
observables of complex systems~\cite{Wang2020}. However, sampling
realizations of stochastic processes is fraught with challenges, particularly when it involves
trajectories that explore rare states.

For instance, in protein folding, although the total folding time may be
of the order of seconds, the time during which the system effectively
jumps from an unfolded state to the folded state can be of
the order of microseconds~\cite{Hartmann2014}. This most interesting part
of the trajectory, during which the system effectively evolves from an
unfolded state to the folded state, is called the transition path. More
precisely, while
the typical time between folding-unfolding events is given by the Kramers
time~\cite{Haenggi1990}, which is exponential in the barrier height (also
known as activation energy), the
duration of the folding process, denoted as the transition-path time
(TPT), is logarithmic in the barrier height~\cite{Zhang2007,Laleman2017}.
This implies that the majority of a simulation is spent waiting for the
rare event of interest to occur and that, to observe a transition, one
might need to run extensive simulations, making the task daunting and
resource intensive.

Langevin bridges have been proposed as an elegant solution to address such
challenges in the case of stochastic
processes of a continuous variable~\cite{Orland2011,Delarue2017,Elber2020,Koehl2022}. They deal
with the problem of sampling the stochastic trajectories of a system,
which starts in a certain known configuration at the initial time and
transitions to a known final state (or family of final states) in a given
time $t_\text{f}$. The goal is to sample many such transition
trajectories~\cite{Neupane2012,Chung2018}, as they allow for the exploration
of the dynamics of the system during the TPT, enabling
the monitoring of large conformational changes undergone by the system.
This is crucial for a microscopic determination of the transition states
and the barrier height of the transition, knowledge pivotal for
applications such as drug design, where modifying the barrier height or
blocking the transition by binding to the transition state is of paramount
importance~\cite{Faccioli2006,Autieri2009,Faccioli2010}.

In this article, we consider the
counterpart of Langevin bridges for stochastic processes of a discrete
variable, namely Markov bridges. While Langevin bridges are derived from a
Fokker-Planck equation and, therefore, require that a force field be given,
we derive Markov bridges from a master equation, which requires that
transition rates between discrete states
be given. The main difficulty in the practical implementation of this
method resides in the efficient evaluation of a
matrix exponential. Yet we will show through several examples that this is
hardly a problem for most applications. In particular, in the context of developmental biology, we will use Markov
bridges to investigate the dynamics of cell
differentiation.

A longstanding challenge in developmental biology is to reconstruct the
temporal sequence of cell differentiation (\emph{i.e.}, cell fate), to understand
commitment to competing cell types, and to understand how robust
differentiation pathways are to fluctuations. Such
questions have traditionally been studied experimentally with
spatio-temporal microscopy data~\cite{McDole2018,Lange2023}, but in recent
years single-cell
RNA sequencing (\scRNA-seq) has emerged as a pivotal technique giving molecular information~\cite{Lange2023} about
cell differentiation. The meticulous examination of RNA abundance within individual cells
has yielded insight into the diversity of cell states in a sample with remarkable quantitative
accuracy, sensitivity, and
throughput~\cite{Sandberg2014,Gawad2016,Schaum2018,Jones2022}. More
recently, the calculation of RNA velocity~\cite{LaManno2018,Bergen2020}, which
estimates the time derivative of RNA abundance in a
single cell for a given gene, has become instrumental in analyzing
dynamic developmental processes involving cell-fate decisions such as embryogenesis and organogenesis.

Computational tools like scVelo~\cite{Bergen2020} and Cell
Rank~\cite{Lange2022,Weiler2023} have been introduced to
compute effective transition rates between cell microstates from RNA
velocities. In this
context, a microstate corresponds to the gene-expression profile of one individual cell. By applying such methods, one
can turn \scRNA developmental data into a discrete Markov chain~\cite{Lange2022} and investigate cell-fate dynamics through the analysis of trajectories of a
variable taking discrete values, namely the microstates.


This article is organized as follows. In~\cref{sec:model} we introduce our
framework and we derive a modified Gillespie
algorithm to sample Markov bridges. In~\cref{sec:results} we show several
applications of our method. Specifically, in~\cref{sec:validation} we validate our method by
applying it to the case of one-dimensional (1D) diffusion, for which analytical
results are available. In~\cref{sec:mueller} we illustrate the benefits of
Markov bridges on a widely used
benchmark potential. Finally, in~\cref{sec:cellfate} we apply our method to the investigation of
the cell-fate dynamics of pancreatic endocrine-cell precursors. We conclude
by discussing the strengths and limitations of our approach
in~\cref{sec:discussion}.

\section{Model}
\label{sec:model}
\subsection{Master equation of a Markov chain}

Let us consider a system with $N$ discrete states, namely $\llbracket 1,
N \rrbracket = \lbrace 1, 2, \ldots, N \rbrace$.
For any pair of states $(\alpha,\beta) \in \llbracket 1, N \rrbracket^2$, we
introduce the transition rate $W_{\alpha \beta}$, such that the
probability to jump from state $\beta$ to state $\alpha$ during an infinitesimal
time interval is $\P \left( \beta, t \to \alpha, t + \dd t \right) =
W_{\alpha \beta} \dd t$. The time evolution of the
probability to be in state $\alpha$ at time $t$ conditioned to the
the initial state $\alpha_\text{i}$ at $t=0$, namely $P_\alpha(t)
= \P \left( \alpha, t \mid \alpha_\text{i}, 0 \right)$ is described by the
master equation~\cite{1992vanKampen_stochastic}:
\begin{align}
  \frac{\dd P_\alpha(t)}{\dd t} = \sum \limits_\beta \left( W_{\alpha \beta}
  P_\beta(t) - W_{\beta \alpha} P_{\alpha}(t)\right).
  \label{eq:master_equation}
\end{align}

Let us now introduce the probability to be in the final
state $\alpha_\text{f}$ at time $t_\text{f}$ conditioned to being in state $\alpha$ at
time $t$, namely $Q_\alpha(t) = \P  \left(\alpha_\text{f}, t_\text{f}
\mid \alpha, t \right)$. By differentiating with respect to $t$ the
identity: $\sum_\alpha P_\alpha(t) Q_\alpha(t) = \P (\alpha_\text{f}, t_\text{f}
| \alpha_\text{i}, 0)$, we obtain the backward master equation:
\begin{align}
  \frac{\dd Q_\alpha(t)}{\dd t} = \sum \limits_\beta W_{\beta \alpha} \left(
  Q_\alpha(t) - Q_{\beta}(t)\right).
  \label{eq:master_equation_backward}
\end{align}

Both~\cref{eq:master_equation,eq:master_equation_backward} can be solved,
and the formal solutions are given by:
\begin{align}
  \begin{aligned}
    P(t) &=  \ee^{t \W} P(0), \\
    Q(t) &=  \ee^{(t_\text{f} -t) \W^T} Q(t_\text{f}),
  \end{aligned}
  \label{eq:master_equation_PQ}
\end{align}
where we introduced the $\W$ matrix~\cite{1992vanKampen_stochastic} with
entries:
\begin{align}
  \mathbbm{W}_{\alpha \beta} = W_{\alpha \beta} - \delta_{\alpha \beta}
  \sum \limits_\gamma W_{\gamma \alpha}.
\end{align}

\subsection{Bridge master equation}
We aim to generate bridges between states $(\alpha_\text{i}, 0)$ and $(\alpha_\text{f},
t_\text{f})$. We therefore express the conditional probability to be in
state $(\alpha, t)$ given the aforementioned initial and final states:
\begin{align}
  \begin{aligned}
    R_\alpha(t) &= \frac{ \P (\alpha_\text{f}, t_\text{f} \mid \alpha,
    t)\P (\alpha, t \mid \alpha_\text{i}, 0)}{\P (\alpha_\text{f}, t_\text{f} \mid
    \alpha_\text{i}, 0)}, \\
    &= \frac{1}{Z} P_{\alpha}(t) Q_\alpha(t),
  \end{aligned}
\end{align}
where $Z$ is a normalization factor ensuring that $\sum_\alpha R_\alpha(t) = 1$. Differentiating the previous
equation, and using~\cref{eq:master_equation,eq:master_equation_backward},
we obtain the master equation satisfied by $R_\alpha(t)$:
\begin{align}
  \begin{aligned}
    \frac{\dd R_\alpha(t)}{\dd t} &= \sum \limits_\beta \left( V_{\alpha
    \beta}(t)
    R_\beta(t) - V_{\beta \alpha}(t) R_{\alpha}(t)\right), \\
    V_{\alpha \beta}(t) &= W_{\alpha \beta} \frac{Q_\alpha(t)}{Q_\beta(t)}.
   \end{aligned}
  \label{eq:master_equation_bridge}
\end{align}

\Cref{eq:master_equation_bridge} is very similar to the master equation in
\cref{eq:master_equation}, except that the original transition rates are
scaled by a time-dependent factor $Q_{\alpha}(t)/Q_\beta(t)$.  When $t_\text{f} -
t \gg 1$, $Q_\alpha(t) \to P_{\alpha_\text{f}}(\infty) = \pi_{\alpha_\text{f}}$ (the
stationary distribution), and we recover the
non-bridge transition rates $V_{\alpha \beta}(t) \to W_{\alpha \beta}$. On
the other hand, when $t \to t_\text{f}$, we have $Q_\alpha(t) \to \delta_{\alpha
\alpha_\text{f}}$, thus $\forall \alpha \neq \alpha_\text{f}$ the incoming transition
rate vanishes, $V_{\alpha \alpha_\text{f}}(t) \to 0$, while the transition rate
towards the final state diverges, $V_{\alpha_\text{f} \alpha} \to +
\infty$. In between the aforementioned limiting cases, the
evaluation of the bridge transition rates requires knowledge of
$Q(t)$.

\subsection{Kinetic Monte Carlo implementation}

To sample bridge trajectories, we can adapt the original
Gillespie algorithm~\cite{Gillespie1976} to the
master~\cref{eq:master_equation_bridge} with time-dependent transition rates.
Given a current state $(\alpha, t)$, the probability to leave this
state in the infinitesimal time interval $[t, t+ \dd t]$ is:
\begin{align}
  \Gamma_\alpha(t) \dd t = \sum \limits_{\beta \ne \alpha} V_{\beta
  \alpha}(t) \dd t.
  \label{eq:escape_rate}
\end{align}

It is convenient to introduce a random variable $T$ representing the dwell
time in state $\alpha$. We define the characteristic function
$F_t(\tau) = \P \left( T < \tau \right)$ and the survival probability
$G_t(\tau) = 1 - F_t(\tau)$. The latter function satisfies the ODE:
\begin{align}
  \frac{\dd G_t}{\dd \tau}(\tau) = - \Gamma_\alpha (t+\tau) G_t(\tau), \quad G_t(0)
  = 1.
\end{align}

The distribution of $T$ is therefore determined:
\begin{align}
  F_t(\tau) = 1 - \exp{\left( - \int \limits_0^\tau \ud{s} \Gamma_\alpha(t + s) \right)}.
\end{align}

By noting that $F_t(T)$ is a uniform random variable between 0 and 1, we can generate $T$ through the operation $T = F_t^{-1}(U)$ where $U \equiv
\mathcal{U}(0,1)$. In practice, $F_t^{-1}$ is not tractable, therefore we
solve for $\tau$ given a realization $u$:
\begin{align}
  \int \limits_0^\tau \ud{s} \Gamma_\alpha(t + s) +
  \ln{(1-u)} = 0.
    \label{eq:dwell_time_draw}
\end{align}

The left-hand side of~\cref{eq:dwell_time_draw} is strictly increasing
with $\tau$ so its root can be easily
found with $0 < \tau < t_\text{f} -t$. Since $V_{\alpha_\text{f} \beta}(t)$ diverges as
$t \to t_\text{f}$, $\Gamma_\alpha(t)$ also diverges as $t \to t_\text{f}$. This ensures
that there is always a solution. Eventually, we can sample Markov
bridges using a modified version of the Gillespie algorithm in which jump
events are determined by solving~\cref{eq:dwell_time_draw}
(see~\cref{sec:pseudocode,al:gillespie} for a pseudo-code implementation).

So far we have omitted giving any explanation on how to compute the
time-dependent transition rates $V_{\alpha \beta}(t)$, yet they are the defining parameters of the bridge master equation,~\cref{eq:master_equation_bridge}. They require
evaluating $Q(t)$, whose formal solution
given in~\cref{eq:master_equation_PQ} relies on the evaluation of a matrix
exponential, which can be computationally expensive for large $N$. In this
work, we decided to use the eigenvalue decomposition of $\W$ to evaluate
this matrix exponential. While costly to compute, the eigenvalue
decomposition $\W = -U \mathrm{Diag}\left(\lambda_1, \ldots, \lambda_N
\right) U^{-1}$ ($0 = \lambda_1 < \lambda_2 \le \ldots \le \lambda_N$) is
computed only once and used subsequently every time $Q(t)$ needs to be
evaluated, as shown in~\cref{eq:qeval}. The Perron-Frobenius theorem
ensures that $\lambda_1 = 0$ is a unique eigenvalue, and that the
associated eigenvector is non-negative: $\pi_\alpha = U_{\alpha 1} \ge 0$.
The vector $\pi$ is the stationary distribution of the Markov process
represented in~\cref{eq:master_equation}. In general, $W$ is a
non-symmetric matrix with real non-negative entries, and the non-symmetric
eigenvalue problem must be solved. When detailed balance is
satisfied, namely $W_{\alpha \beta} \pi_\beta = W_{\beta \alpha} \pi_\alpha$, the matrix
$\pi_\alpha^{-1/2} W_{\alpha \beta} \pi_\beta^{1/2}$ is symmetric and can therefore be
diagonalized efficiently, along with $\W$. For large
systems, the eigenvalue decomposition of $\W$ (and most other matrix decompositions
of the form $A = U B U^{-1}$) might be difficult to obtain  since the usual methods require $\oO{N^3}$ computing time. For
such systems, one can still sample Markov bridges using the standard
method to sample Markov chains (see~\cref{sec:pseudocode}).

\begin{align}
  Q_\alpha(t) = \sum \limits_k U_{\alpha_\text{f} k} \left(U^{-1}\right)_{k \alpha}
  \ee^{-\lambda_k (t_\text{f} - t)}.
  \label{eq:qeval}
\end{align}

In this work, we have restrained ourselves to Markov bridges connecting
one initial state $(\alpha_\text{i}, 0)$ to one final state $(\alpha_\text{f}, t_\text{f})$.
Yet, our method can easily be extended to Markov bridges with $L$
competing final states $\lbrace \alpha_\text{f}^{(l)} \rbrace$. For example, to
generate bridges with equal probabilities to end in any of the competing
final states~\cref{eq:master_equation_PQ} needs to be solved with:
\begin{align}
  Q_\alpha(t_\text{f}) = \frac{1}{L} \sum \limits_{l = 1}^L \delta_{\alpha \alpha_\text{f}^{(l)}},
\end{align}
and~\cref{eq:qeval} must be modified to:
\begin{align}
  Q_\alpha(t) = \frac{1}{L} \sum \limits_{l=1}^L \sum \limits_k U_{\alpha_\text{f}^{(l)} k} \left(U^{-1}\right)_{k \alpha} \ee^{-\lambda_k (t_\text{f} - t)}.
\end{align}

\section{Results}
\label{sec:results}

\begin{figDiff}
  \centering
  \includegraphics[width = 3.5 in, page=1]{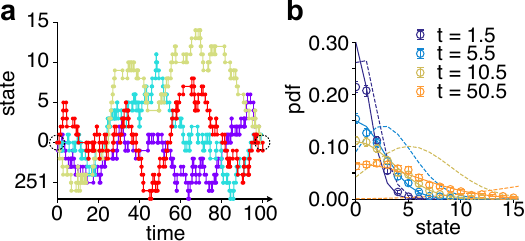}
  \caption{
    Markov bridges for diffusion on the 1D lattice, with
   $N=256$, $\alpha = 1$, $\beta = 0.5$, $n_\text{i} = n_\text{f} = 0$ and $t_\text{f} = 100$.
   \textbf{(a)} Four selected trajectories. \textbf{(b)} The computed
   estimate for $R(t)$ (symbols) agrees with the theoretical expression (solid
  lines), but is markedly different from $P(t)$ that would be obtained
  with the non-bridge dynamics (dashed lines). We show the distributions at times
  $t=1.5,\, 5.5,\, 10.5,\, 50.5$.}
  \label{fig:diffusion}
\end{figDiff}

We now proceed to apply our method to several examples. We start with 1D
diffusion on a lattice. Although trivial, since $P(t)$, $Q(t)$ and $R(t)$
can be computed exactly, it allows us to validate our model, by comparing
the results of the simulations to the theory. Second, we apply our method
to sample bridge trajectories to study the transition between two wells in
the M\"{u}ller-Brown potential, which is widely used in physical chemistry
as a benchmark potential to investigate transition paths. Theoretical
results are not available, yet this example will help us illustrate the
benefits of our method which enables us to focus on the transition path and
to discard the less interesting part of the simulation spent waiting for
the transition to occur. Third, we illustrate the interest of our method
in studying cell fate. Recent developments have made it possible to
estimate transition rates from one cell type to another from \scRNA-seq data. This is achieved by computing so-called RNA
velocities, which quantify the rate of change in the genes' expression
profiles. We will generate Markov bridges for two differentiation pathways
of pancreatic endocrine-cell precursors.

\subsection{Validation with 1D diffusion}
\label{sec:validation}

We first set out to validate our method by applying it to diffusion on the 1D
lattice, for which
$P(t)$ and $Q(t)$ can be expressed analytically. For this process,
the master equation from~\cref{eq:master_equation} reads:
\begin{align}
  \frac{\dd P_n}{\dd t} = \alpha P_{n-1} + \beta P_{n+1} - (\alpha +
  \beta) P_n, \qquad \forall n \in \llbracket 1, N \rrbracket,
  \label{eq:diffusion_mequation_P}
\end{align}
where $n \in \llbracket 1, N \rrbracket$ denotes the state of
a diffusing particle being at position $x_n = n \Delta x$ ($\Delta x$ is the lattice
site size), and $\alpha$ (respectively $\beta$) is the forward
(respectively backward) transition rate. The diffusion
process is biased when $\alpha \neq \beta$. We consider periodic boundary
conditions, \textit{i.e.}, $x_{N+1} = x_1$.
We generated \num{1024} trajectories using~\cref{al:gillespie}, for which we used
the following parameters: $N=256$, $\alpha = 1$, $\beta = \num{0.5}$
(hence the diffusion is biased), $n_\text{i} = n_\text{f} = 0$ and $t_\text{f} = 100$.
In~\cref{fig:diffusion}a, we show 4 of those trajectories.

Since $P(t)$ and $Q(t)$ can be expressed analytically, so can the
bridge probability distribution $R(t)$ be (see \cref{sec:diffusion}). We can
therefore compare this theoretical distribution to its estimate obtained
by averaging over the sampled trajectories. For this purpose we defined a
time subdivision $t_j = j \Delta t $ for $j \in \llbracket 0, M \rrbracket$,
such that $t_M = t_\text{f}$. The estimate of $R(t_j)$ from simulations is then:
\begin{align}
  R_n^\mathrm{sim}(t_j) = \frac{1}{Z_j} \sum \limits_s \sum
  \limits_{(n_s,t_s)} \delta_{n_s, n} \mathbbm{1}_{[t_j, t_{j+1}[}(t_s),
\end{align}
where the sum runs over all states $(n_s, t_s)$ sampled in each
bridge trajectory with index $s$, and $Z_j$ is a normalization factor
ensuring that $\sum_n R_n^\mathrm{sim}(t_j) = 1$.
In~\cref{fig:diffusion}b, we computed $R^\mathrm{sim}$ at $t_j = 1.5,\,
5,5,\, 10.5,\, 50.5$, with a subdivision size $\Delta t = 1$. We find that
it agrees with the theoretical values of $R(t)$ (solid lines). For
comparison, we also show the theoretical values of $P(t)$ (dashed
lines), which are drastically different. Therefore the method used to
sample Markov bridges successfully recovers expected theoretical results.

\subsection{Application to the M\"{u}ller-Brown potential}
\label{sec:mueller}

The M\"{u}ller-Brown potential~\cite{Mueller1979,Mueller1980} is a standard
benchmark potential used to check the validity of methods for generating
transition paths. It is a two-dimensional potential given by:
\begin{align}
  U(x,y) = \sum \limits_{j=1}^4 A_j \ee^{a_j(x-x_j^0)^2 +
  b_j(x-x_j^0)(y-y_j^0) + c_j (y-y_j^0)^2},
\end{align}
with:
\begin{align}
  \begin{aligned}
    A &= [-200, -100, -170, 15], \\
    a &= [-1, -1, -6.5, 0.7], \\
    b &= [0, 0, 11, 0.6], \\
    c &= [-10, -10, -6.5, 0.7], \\
    x^0 &= [1, 0, -0.5, -1], \\
    y^0 &= [0, 0.5, 1.5, 1].
  \end{aligned}
\end{align}

It has 3 minima denoted by A $(-0.558, 1.442)$, B $(-0.05, 0.467)$ and C
$(0.623, 0.028)$. As a second application of our method, we seek to sample
Markov bridges from A to C. For that matter, we considered a 2D grid of points in
$[-1.5,1] \times [-0.5, 2]$, which defines $N=4225$ states. We defined the
transition rates as given in~\cref{eq:rates_mb}. We note that detailed balance is satisfied~\cite{Binder1992},
namely $W_{\alpha \beta} \pi_\beta = W_{\beta \alpha} \pi_\alpha$. This
ensures that the stationary distribution is the Boltzmann distribution:
$\pi_\alpha \propto \exp{(-U_\alpha / k_\mathrm{B} T)}$.
\begin{align}
  W_{\alpha \beta} = \ee^{-\frac{U_\alpha - U_{\beta}}{2 k_\mathrm{B} T}}.
  \label{eq:rates_mb}
\end{align}

\begin{figMuellerApplication}
  \centering
  \includegraphics[width = 3.5 in, page=2]{figures.pdf}
  \caption{
    \textbf{(a)} Markov bridges for the M\"{u}ller-Brown potential, with
   $N=4225$, $\alpha_\text{i} = A$, $\alpha_\text{f} = C$ and $t_\text{f} = 1000$. The 2D grid
   of microstates is represented with black crosses ($+$). The level lines
  of the M\"{u}ller-Brown potential are shown (thin solid black lines). We also give
   the mean trajectory (thick black line) and the most likely trajectory
   (thick red line). We show a selection of trajectories going through
   \textbf{(b)} low and \textbf{(c)} high
   energy barriers.}
  \label{fig:mb}
\end{figMuellerApplication}

To accommodate the accuracy of double precision computation arithmetic,
we truncated the potential so that $\min U(x,y) = U_\mathrm{min} = -75 \, k_\mathrm{B} T$
and $\max U(x,y) = U_\mathrm{max} = 15 \, k_\mathrm{B} T$, and we chose
$k_\mathrm{B} T =
3.2$, so that the ratio of probabilities of the extreme states is $\exp{\left( \left(U_\mathrm{max}-U_\mathrm{min}\right) /
k_\mathrm{B} T \right)} \approx
10^{-15}$. In~\cref{fig:mb} we show \num{128} Markov bridges generated with our
method. We also represented the mean and most likely trajectory
(details in \cref{sec:mean_mostl_traj}). As can be expected most
trajectories take paths close to the most likely trajectory
(\cref{fig:mb}b). Yet some of
them use paths which go through high energy barriers (\cref{fig:mb}c).
The mean trajectory is markedly different from the most likely
trajectory. This is mostly due to the inhomogeneous kinetics of the
bridge trajectories. Indeed, each trajectory spends a varying amount of
time sampling states in the potential well $A$ before
actually committing to transitioning to towards states $B$ and $C$. This
asynchrony in the timing of the transition $A \to C$ causes the mean
trajectory to deviate from the most likely one. For the same reason, the
most likely trajectory erroneously suggests a small number of large jumps
connecting the wells surrounding states $A$, $B$ and $C$.

\begin{figMuellerTP}
  \centering
  \includegraphics[width = 3.5 in, page=3]{figures.pdf}
  \caption{
    Histograms for the \textbf{(a)} first-passage times and \textbf{(b)} transition-path
    times obtained by sampling \num{2048} trajectories with standard \KMC
    (\cref{eq:master_equation}) in the M\"{u}ller-Brown potential, with
   $N=4225$ and $\alpha_\text{i} = A$. \textbf{(c)} Transition paths for a
   selection of the most-likely trajectories (left) and trajectories
   visiting the rarest states (right). The colors indicate the states
   belonging to either of the 3 wells in the M\"{u}ller-Brown potential.
   Filled circle indicates the bottom of each well ($U(x,y) = U_A, U_B,
   U_C$). The thick dashed line denotes the mean transition path
   (see~\cref{sec:mean_mostl_traj}).
 }
  \label{fig:transition_paths}
\end{figMuellerTP}

This example also emphasizes the difference between first-passage time
(FPT) and TPT. To be more quantitative, we
considered the non-bridge dynamics described
in~\cref{eq:master_equation}, and we sampled \num{2048} trajectories
using the Gillespie algorithm.  For each trajectory, we defined the
last-passage time (LPT) as the time corresponding to the last occurrence of a state belonging to
the bottom of well $A$ (filled circles in~\cref{fig:transition_paths}c),
and the FPT as the time corresponding to the first occurrence of a state
belonging to well $C$. The transition-path time is then simply given by
$\text{TPT} = \text{FPT}-\text{LPT}$. We checked that the
distribution of the TPT agrees with previously reported theoretical
results (\cref{fig:tpt_fit,sec:tpt_pdf}). \Cref{fig:transition_paths}a-b show that
the mean TPT is about 4 orders of magnitude smaller than
the mean FPT. Therefore to study the transition $A
\to C$ with standard \KMC, one has to typically simulate trajectories for a duration
of time of the order of \num{10^4} time units, whereas the portion of
interest in which the transition $A \to C$ actually happens represents
less than \SI{0.01}{\percent} of the trajectory. Computationnally, such an
approach is very inefficient. This emphasizes the benefits of our method,
which allows to specifically sample the transition paths.  That also means
that the value for $t_\text{f}$ should be chosen to be slightly larger than the
mean TPT.

\subsection{Dynamics of differentiation pathways of pancreatic endocrine-cell precursors}
\label{sec:cellfate}

We now use our method to generate cell-fate trajectories. We used a mouse pancreatic development
dataset~\cite{BastidasPonce2019} which comprises $N=3696$
states (see~\cref{sec:pancreas}). Single
cells can be visualized by applying dimensional reduction techniques to
the gene expression vectors. In particular, they can be projected onto
the first two components obtained by the Uniform Manifold Approximation
and Projection method (UMAP)~\cite{McInnes2018,Becht2019}, and we computed
the transition rates using
scVelo (see~\cref{sec:pancreas}). We
noted that the resulting rate matrix $\W$ did not satisfy detailed balance.
Using our method, we generated \num{512} Markov bridges to study the
transition from the ductal cell type to the beta cell type~(\cref{fig:scrna}a) and from
the ductal cell type to the alpha cell type~(\cref{fig:scrna}b). To
illustrate how biological insights can be gained, we focus on the
differentiation into the alpha cell type. The
time-lapse in~\cref{fig:scrna_insights}a shows that trajectories spend a
variable but significant amount of time in the ductal cell type. This is
further corroborated by noting that the average waiting time in the ductal
and Ngn3 low-EP cell types are the longest (\cref{fig:scrna_insights}c).
Here the waiting time was defined as the time of the last occurrence of a
cell type minus the its first occurrence, \emph{i.e.}, $\text{LPT} -
\text{FPT}$.
Therefore, commitment out of the ductal and Ngn3 low-EP cell types
appears to be the main bottleneck of this differentiation route.

\begin{figScrnaApplication}
  \centering
  \includegraphics[width = 3.0 in, page=4]{figures.pdf}
  \caption{
    Markov bridges for the pancreas dataset, with
   $N=3696$, $\alpha_\text{i} = 1103$ (ductal cell type) and $t_\text{f} = 100$. We
   study the transition towards the final state \textbf{(a)} $\alpha_\text{f} =
   2379$ (beta cell type) and \textbf{(b)} $\alpha_\text{f} = 3437$ (alpha cell
   type). We generated $512$ bridge trajectories in each case. The state
   coordinates are the first two components of a UMAP analysis.
 }
  \label{fig:scrna}
\end{figScrnaApplication}

Once commitment to the Ngn3 high-EP cell type occurs, trajectories move
primarily from left to right, with very few jumps in the reverse
direction, as qualitatively seen in the time-lapse
(\cref{fig:scrna_insights}a). For a more precise statement, we computed
for each cell type the mean direction by
averaging the curve tangent (velocity normalized to unity,
$\vec{u} = \vec{v} / \| \vec{v} \|$) over all states in all trajectories
which are in a given cell type. We
obtain that the cell types exhibiting the most robust direction are the
epsilon, Ngn3 high-EP and pre-endocrine cell types
(\cref{fig:scrna_insights}c). While the ductal and Ngn3 low-EP were
previously qualified as bottlenecks, these three states should rather be
seen as the highways for this differentiation route.

Another insight that can be gained from analyzing bridge trajectories
concerns the statistical weights of the different routes for
differentiation. In~\cref{fig:scrna_insights}b we represented only the
part of the trajectories which is between the last occurrence of the Ngn3
high-EP cell type and the first occurrence of the alpha cell type. The
trajectories can be broadly divided into three groups, each taking a
different route to connect those two cell types. It appears that one of
the group is by far the most-likely, containing about \SI{94}{\percent} of
the sampled trajectories. The two other groups have similar weight.
Interestingly, it seems that the epsilon cell type is only accessible
through these less likely routes. We speculate that this approach can be
used to investigate more broadly bifurcations in cell-fate dynamics.

We note that velocities, akin to RNA velocities, can be obtained directly
from the transition rates, as:
\begin{align}
  \vec{v}_\alpha = \sum \limits_{\beta \ne \alpha} (\vec{x}_\beta -
  \vec{x}_\alpha) W_{\beta \alpha},
\end{align}
which seemingly does not require sampling Markov bridges. While the
velocity field obtained (\cref{fig:scrna_velocities}) is consistent with
the observations made above (for example the bottleneck of the ductal/Ngn3
low-EP cell type appears as a region with velocities pointing inward), it
is not enough by itself to get a temporal resolution of the dynamics nor
to analyze the role of fluctuations such as the statistical weights of
different routes, therefore emphasizing the unique type of insights that can be gained by
using Markov bridges to study differentiation pathways.

\begin{figScrnaInsights}
  \centering
  \includegraphics[width = \linewidth, page=5]{figures.pdf}
  \caption{Insights for the differentiation of pancreatic endocrine-cell
  precursors to alpha cells obtained from the bridge trajectories shown
  in~\cref{fig:scrna}. \textbf{(a)} Time-lapse of the bridge dynamics.
  Current states are represented with filled circles and the tails show a
  few of the previous states.
  \textbf{(b)} Focus on the parts of the bridge trajectories which go from
  the ``Ngn3 high-EP'' to ``alpha'' cell types. \textbf{(c)} Norm of the mean
  direction ($\| \vec{u} \|$) and waiting time
  ($\text{LPT} - \text{FPT}$)
  for each cell type. The inset shows the mean direction vector $\vec{u}$
  for the 8 different cell types.
 }
  \label{fig:scrna_insights}
\end{figScrnaInsights}

\section{Discussion}
\label{sec:discussion}

The evaluation of the bridge
transition rates requires the evaluation of $Q(t)$, which, as already
mentioned, involves a matrix exponential (\cref{eq:master_equation_PQ}).
In this work, we have made the choice to use the eigenvalue decomposition
of $\W$ to evaluate $Q(t)$ as shown in~\cref{eq:qeval}. However, in
general, for any matrix $A = U D U^{-1}$ ($D = \mathrm{Diag}\left(
\lambda_1, \lambda_2, \ldots, \lambda_N \right)$), $\ee^{t A}$ is not well
approximated (numerically) by $U \ee^D U^{-1}$ when A is non-normal (see
for example~\cite{Moler2003} and section 11.3 of~\cite{Golub2013}).
Unfortunately, $\W$ is seldom normal. Therefore, an improvement to our
method would be to find more reliable ways to compute $Q(t)$.

Given a system with discrete states, whose dynamics can be described by a
master equation as in~\cref{eq:master_equation}, we have presented a
method to generate the subset of possible trajectories that go from one
initial state $\alpha_\text{i}$ to a final state $\alpha_\text{f}$ in a time $t_\text{f}$. This
method is especially valuable when the final state has a low probability
to occur, or when the initial state and the final state are separated by
high energy barriers (in the sense of Kramers' theory). Stochastic
realizations of such processes might exhibit
a large FPT, and would be sampled with very low efficiency using standard
\KMC methods (e.g. Gillespie algorithm). In such cases, most of the
time is therefore spent waiting for a rare event to happen. Yet, in
most applications, the sampling of states during this waiting period is of
little interest; we are rather interested in the actual transition path once a
trajectory has committed to the transition. A benefit of our method is
therefore to discard the uninteresting waiting time. This can be done
heuristically by adjusting $t_\text{f}$ so that it is slightly greater than the
typical TPT.

Although we mentioned some lines of improvement, we showed
that our approach can be applied in its current state to study diverse
phenomena, including cell-fate dynamics, and yield unique insights. The
application to \scRNA-seq data paves the way for novel computational
approaches to study cell development, such as \textit{in silico}
perturbations. We note that it would be interesting to compare results
from our methods with experiments tracking cell fate with \scRNA-seq and
an orthogonal modality (e.g. microscopy), when they become available.
Finally, in this work a cell microstate was characterized only
through RNA abundance. However it could be enriched with observables
coming from other modalities (e.g. morphological measurements coming from
microscopy imaging). Provided that an appropriate definition of transition
rates is taken, Markov bridges could therefore be used to probe more complex features
of cell-fate dynamics.


\clearpage
\small
\begin{acknowledgments}
\section*{Acknowledgements}
G.L.T. wishes to thank Lo\"ic Royer for very useful discussions, and Alejandro Granados for introducing him to single-cell RNA sequencing and planting the seed for the application of Markov bridges to cell-fate choice.

\end{acknowledgments}

\normalsize
\bibliography{bibliography}

\begin{thebibliography}{39}%
\makeatletter
\providecommand \@ifxundefined [1]{%
 \@ifx{#1\undefined}
}%
\providecommand \@ifnum [1]{%
 \ifnum #1\expandafter \@firstoftwo
 \else \expandafter \@secondoftwo
 \fi
}%
\providecommand \@ifx [1]{%
 \ifx #1\expandafter \@firstoftwo
 \else \expandafter \@secondoftwo
 \fi
}%
\providecommand \natexlab [1]{#1}%
\providecommand \enquote  [1]{``#1''}%
\providecommand \bibnamefont  [1]{#1}%
\providecommand \bibfnamefont [1]{#1}%
\providecommand \citenamefont [1]{#1}%
\providecommand \href@noop [0]{\@secondoftwo}%
\providecommand \href [0]{\begingroup \@sanitize@url \@href}%
\providecommand \@href[1]{\@@startlink{#1}\@@href}%
\providecommand \@@href[1]{\endgroup#1\@@endlink}%
\providecommand \@sanitize@url [0]{\catcode `\\12\catcode `\$12\catcode
  `\&12\catcode `\#12\catcode `\^12\catcode `\_12\catcode `\%12\relax}%
\providecommand \@@startlink[1]{}%
\providecommand \@@endlink[0]{}%
\providecommand \url  [0]{\begingroup\@sanitize@url \@url }%
\providecommand \@url [1]{\endgroup\@href {#1}{\urlprefix }}%
\providecommand \urlprefix  [0]{URL }%
\providecommand \Eprint [0]{\href }%
\providecommand \doibase [0]{https://doi.org/}%
\providecommand \selectlanguage [0]{\@gobble}%
\providecommand \bibinfo  [0]{\@secondoftwo}%
\providecommand \bibfield  [0]{\@secondoftwo}%
\providecommand \translation [1]{[#1]}%
\providecommand \BibitemOpen [0]{}%
\providecommand \bibitemStop [0]{}%
\providecommand \bibitemNoStop [0]{.\EOS\space}%
\providecommand \EOS [0]{\spacefactor3000\relax}%
\providecommand \BibitemShut  [1]{\csname bibitem#1\endcsname}%
\let\auto@bib@innerbib\@empty
\bibitem [{\citenamefont {Shaw}\ \emph {et~al.}(2008)\citenamefont {Shaw},
  \citenamefont {Deneroff}, \citenamefont {Dror}, \citenamefont {Kuskin},
  \citenamefont {Larson}, \citenamefont {Salmon}, \citenamefont {Young},
  \citenamefont {Batson}, \citenamefont {Bowers}, \citenamefont {Chao},
  \citenamefont {Eastwood}, \citenamefont {Gagliardo}, \citenamefont
  {Grossman}, \citenamefont {Ho}, \citenamefont {Ierardi}, \citenamefont
  {Kolossváry}, \citenamefont {Klepeis}, \citenamefont {Layman}, \citenamefont
  {McLeavey}, \citenamefont {Moraes}, \citenamefont {Mueller}, \citenamefont
  {Priest}, \citenamefont {Shan}, \citenamefont {Spengler}, \citenamefont
  {Theobald}, \citenamefont {Towles},\ and\ \citenamefont {Wang}}]{Shaw2008}%
  \BibitemOpen
  \bibfield  {author} {\bibinfo {author} {\bibfnamefont {D.~E.}\ \bibnamefont
  {Shaw}}, \bibinfo {author} {\bibfnamefont {M.~M.}\ \bibnamefont {Deneroff}},
  \bibinfo {author} {\bibfnamefont {R.~O.}\ \bibnamefont {Dror}}, \bibinfo
  {author} {\bibfnamefont {J.~S.}\ \bibnamefont {Kuskin}}, \bibinfo {author}
  {\bibfnamefont {R.~H.}\ \bibnamefont {Larson}}, \bibinfo {author}
  {\bibfnamefont {J.~K.}\ \bibnamefont {Salmon}}, \bibinfo {author}
  {\bibfnamefont {C.}~\bibnamefont {Young}}, \bibinfo {author} {\bibfnamefont
  {B.}~\bibnamefont {Batson}}, \bibinfo {author} {\bibfnamefont {K.~J.}\
  \bibnamefont {Bowers}}, \bibinfo {author} {\bibfnamefont {J.~C.}\
  \bibnamefont {Chao}}, \bibinfo {author} {\bibfnamefont {M.~P.}\ \bibnamefont
  {Eastwood}}, \bibinfo {author} {\bibfnamefont {J.}~\bibnamefont {Gagliardo}},
  \bibinfo {author} {\bibfnamefont {J.~P.}\ \bibnamefont {Grossman}}, \bibinfo
  {author} {\bibfnamefont {C.~R.}\ \bibnamefont {Ho}}, \bibinfo {author}
  {\bibfnamefont {D.~J.}\ \bibnamefont {Ierardi}}, \bibinfo {author}
  {\bibfnamefont {I.}~\bibnamefont {Kolossváry}}, \bibinfo {author}
  {\bibfnamefont {J.~L.}\ \bibnamefont {Klepeis}}, \bibinfo {author}
  {\bibfnamefont {T.}~\bibnamefont {Layman}}, \bibinfo {author} {\bibfnamefont
  {C.}~\bibnamefont {McLeavey}}, \bibinfo {author} {\bibfnamefont {M.~A.}\
  \bibnamefont {Moraes}}, \bibinfo {author} {\bibfnamefont {R.}~\bibnamefont
  {Mueller}}, \bibinfo {author} {\bibfnamefont {E.~C.}\ \bibnamefont {Priest}},
  \bibinfo {author} {\bibfnamefont {Y.}~\bibnamefont {Shan}}, \bibinfo {author}
  {\bibfnamefont {J.}~\bibnamefont {Spengler}}, \bibinfo {author}
  {\bibfnamefont {M.}~\bibnamefont {Theobald}}, \bibinfo {author}
  {\bibfnamefont {B.}~\bibnamefont {Towles}},\ and\ \bibinfo {author}
  {\bibfnamefont {S.~C.}\ \bibnamefont {Wang}},\ }\bibfield  {title} {\bibinfo
  {title} {Anton, a special-purpose machine for molecular dynamics
  simulation},\ }\href {https://doi.org/10.1145/1364782.1364802} {\bibfield
  {journal} {\bibinfo  {journal} {Communications of the ACM}\ }\textbf
  {\bibinfo {volume} {51}},\ \bibinfo {pages} {91} (\bibinfo {year}
  {2008})}\BibitemShut {NoStop}%
\bibitem [{\citenamefont {Elber}(2011)}]{Elber2011}%
  \BibitemOpen
  \bibfield  {author} {\bibinfo {author} {\bibfnamefont {R.}~\bibnamefont
  {Elber}},\ }\bibfield  {title} {\bibinfo {title} {Simulations of allosteric
  transitions},\ }\href {https://doi.org/10.1016/j.sbi.2011.01.012} {\bibfield
  {journal} {\bibinfo  {journal} {Current Opinion in Structural Biology}\
  }\textbf {\bibinfo {volume} {21}},\ \bibinfo {pages} {167} (\bibinfo {year}
  {2011})}\BibitemShut {NoStop}%
\bibitem [{\citenamefont {Cavagna}(2009)}]{Cavagna2009}%
  \BibitemOpen
  \bibfield  {author} {\bibinfo {author} {\bibfnamefont {A.}~\bibnamefont
  {Cavagna}},\ }\bibfield  {title} {\bibinfo {title} {Supercooled liquids for
  pedestrians},\ }\href {https://doi.org/10.1016/j.physrep.2009.03.003}
  {\bibfield  {journal} {\bibinfo  {journal} {Physics Reports}\ }\textbf
  {\bibinfo {volume} {476}},\ \bibinfo {pages} {51} (\bibinfo {year}
  {2009})}\BibitemShut {NoStop}%
\bibitem [{\citenamefont {Charbonneau}\ \emph {et~al.}(2023)\citenamefont
  {Charbonneau}, \citenamefont {Marinari}, \citenamefont {Parisi},
  \citenamefont {Ricci-tersenghi}, \citenamefont {Sicuro}, \citenamefont
  {Zamponi},\ and\ \citenamefont {Mezard}}]{Charbonneau2023}%
  \BibitemOpen
  \bibfield  {author} {\bibinfo {author} {\bibfnamefont {P.}~\bibnamefont
  {Charbonneau}}, \bibinfo {author} {\bibfnamefont {E.}~\bibnamefont
  {Marinari}}, \bibinfo {author} {\bibfnamefont {G.}~\bibnamefont {Parisi}},
  \bibinfo {author} {\bibfnamefont {F.}~\bibnamefont {Ricci-tersenghi}},
  \bibinfo {author} {\bibfnamefont {G.}~\bibnamefont {Sicuro}}, \bibinfo
  {author} {\bibfnamefont {F.}~\bibnamefont {Zamponi}},\ and\ \bibinfo {author}
  {\bibfnamefont {M.}~\bibnamefont {Mezard}},\ }\href@noop {} {\emph {\bibinfo
  {title} {Spin Glass Theory and Far Beyond: Replica Symmetry Breaking after 40
  Years}}}\ (\bibinfo  {publisher} {World Scientific},\ \bibinfo {year}
  {2023})\BibitemShut {NoStop}%
\bibitem [{\citenamefont {Shan}\ \emph {et~al.}(2011)\citenamefont {Shan},
  \citenamefont {Kim}, \citenamefont {Eastwood}, \citenamefont {Dror},
  \citenamefont {Seeliger},\ and\ \citenamefont {Shaw}}]{Shan2011}%
  \BibitemOpen
  \bibfield  {author} {\bibinfo {author} {\bibfnamefont {Y.}~\bibnamefont
  {Shan}}, \bibinfo {author} {\bibfnamefont {E.~T.}\ \bibnamefont {Kim}},
  \bibinfo {author} {\bibfnamefont {M.~P.}\ \bibnamefont {Eastwood}}, \bibinfo
  {author} {\bibfnamefont {R.~O.}\ \bibnamefont {Dror}}, \bibinfo {author}
  {\bibfnamefont {M.~A.}\ \bibnamefont {Seeliger}},\ and\ \bibinfo {author}
  {\bibfnamefont {D.~E.}\ \bibnamefont {Shaw}},\ }\bibfield  {title} {\bibinfo
  {title} {How {Does} a {Drug} {Molecule} {Find} {Its} {Target} {Binding}
  {Site}?},\ }\href {https://doi.org/10.1021/ja202726y} {\bibfield  {journal}
  {\bibinfo  {journal} {Journal of the American Chemical Society}\ }\textbf
  {\bibinfo {volume} {133}},\ \bibinfo {pages} {9181} (\bibinfo {year}
  {2011})}\BibitemShut {NoStop}%
\bibitem [{\citenamefont {Wang}\ \emph {et~al.}(2020)\citenamefont {Wang},
  \citenamefont {Ramírez-Hinestrosa},\ and\ \citenamefont
  {Frenkel}}]{Wang2020}%
  \BibitemOpen
  \bibfield  {author} {\bibinfo {author} {\bibfnamefont {X.}~\bibnamefont
  {Wang}}, \bibinfo {author} {\bibfnamefont {S.}~\bibnamefont
  {Ramírez-Hinestrosa}},\ and\ \bibinfo {author} {\bibfnamefont
  {D.}~\bibnamefont {Frenkel}},\ }\bibfield  {title} {\bibinfo {title} {Using
  {Molecular} {Simulation} to {Compute} {Transport} {Coefficients} of
  {Molecular} {Gases}},\ }\href {https://doi.org/10.1021/acs.jpcb.0c04462}
  {\bibfield  {journal} {\bibinfo  {journal} {The Journal of Physical Chemistry
  B}\ }\textbf {\bibinfo {volume} {124}},\ \bibinfo {pages} {7636} (\bibinfo
  {year} {2020})}\BibitemShut {NoStop}%
\bibitem [{\citenamefont {Hartmann}\ \emph {et~al.}(2014)\citenamefont
  {Hartmann}, \citenamefont {Banisch}, \citenamefont {Sarich}, \citenamefont
  {Badowski},\ and\ \citenamefont {Sch\"{u}tte}}]{Hartmann2014}%
  \BibitemOpen
  \bibfield  {author} {\bibinfo {author} {\bibfnamefont {C.}~\bibnamefont
  {Hartmann}}, \bibinfo {author} {\bibfnamefont {R.}~\bibnamefont {Banisch}},
  \bibinfo {author} {\bibfnamefont {M.}~\bibnamefont {Sarich}}, \bibinfo
  {author} {\bibfnamefont {T.}~\bibnamefont {Badowski}},\ and\ \bibinfo
  {author} {\bibfnamefont {C.}~\bibnamefont {Sch\"{u}tte}},\ }\bibfield
  {title} {\bibinfo {title} {Characterization of {Rare} {Events} in {Molecular}
  {Dynamics}},\ }\href {https://doi.org/10.3390/e16010350} {\bibfield
  {journal} {\bibinfo  {journal} {Entropy}\ }\textbf {\bibinfo {volume} {16}},\
  \bibinfo {pages} {350} (\bibinfo {year} {2014})}\BibitemShut {NoStop}%
\bibitem [{\citenamefont {H\"{a}nggi}\ \emph {et~al.}(1990)\citenamefont
  {H\"{a}nggi}, \citenamefont {Talkner},\ and\ \citenamefont
  {Borkovec}}]{Haenggi1990}%
  \BibitemOpen
  \bibfield  {author} {\bibinfo {author} {\bibfnamefont {P.}~\bibnamefont
  {H\"{a}nggi}}, \bibinfo {author} {\bibfnamefont {P.}~\bibnamefont
  {Talkner}},\ and\ \bibinfo {author} {\bibfnamefont {M.}~\bibnamefont
  {Borkovec}},\ }\bibfield  {title} {\bibinfo {title} {Reaction-rate theory:
  fifty years after {Kramers}},\ }\href
  {https://doi.org/10.1103/RevModPhys.62.251} {\bibfield  {journal} {\bibinfo
  {journal} {Reviews of Modern Physics}\ }\textbf {\bibinfo {volume} {62}},\
  \bibinfo {pages} {251} (\bibinfo {year} {1990})}\BibitemShut {NoStop}%
\bibitem [{\citenamefont {Zhang}\ \emph {et~al.}(2007)\citenamefont {Zhang},
  \citenamefont {Jasnow},\ and\ \citenamefont {Zuckerman}}]{Zhang2007}%
  \BibitemOpen
  \bibfield  {author} {\bibinfo {author} {\bibfnamefont {B.~W.}\ \bibnamefont
  {Zhang}}, \bibinfo {author} {\bibfnamefont {D.}~\bibnamefont {Jasnow}},\ and\
  \bibinfo {author} {\bibfnamefont {D.~M.}\ \bibnamefont {Zuckerman}},\
  }\bibfield  {title} {\bibinfo {title} {Transition-event durations in
  one-dimensional activated processes},\ }\href
  {https://doi.org/10.1063/1.2434966} {\bibfield  {journal} {\bibinfo
  {journal} {The Journal of Chemical Physics}\ }\textbf {\bibinfo {volume}
  {126}},\ \bibinfo {pages} {074504} (\bibinfo {year} {2007})}\BibitemShut
  {NoStop}%
\bibitem [{\citenamefont {Laleman}\ \emph {et~al.}(2017)\citenamefont
  {Laleman}, \citenamefont {Carlon},\ and\ \citenamefont
  {Orland}}]{Laleman2017}%
  \BibitemOpen
  \bibfield  {author} {\bibinfo {author} {\bibfnamefont {M.}~\bibnamefont
  {Laleman}}, \bibinfo {author} {\bibfnamefont {E.}~\bibnamefont {Carlon}},\
  and\ \bibinfo {author} {\bibfnamefont {H.}~\bibnamefont {Orland}},\
  }\bibfield  {title} {\bibinfo {title} {Transition path time distributions},\
  }\href {https://doi.org/10.1063/1.5000423} {\bibfield  {journal} {\bibinfo
  {journal} {The Journal of Chemical Physics}\ }\textbf {\bibinfo {volume}
  {147}},\ \bibinfo {pages} {214103} (\bibinfo {year} {2017})}\BibitemShut
  {NoStop}%
\bibitem [{\citenamefont {Orland}(2011)}]{Orland2011}%
  \BibitemOpen
  \bibfield  {author} {\bibinfo {author} {\bibfnamefont {H.}~\bibnamefont
  {Orland}},\ }\bibfield  {title} {\bibinfo {title} {Generating transition
  paths by {Langevin} bridges},\ }\href {https://doi.org/10.1063/1.3586036}
  {\bibfield  {journal} {\bibinfo  {journal} {The Journal of Chemical Physics}\
  }\textbf {\bibinfo {volume} {134}},\ \bibinfo {pages} {174114} (\bibinfo
  {year} {2011})}\BibitemShut {NoStop}%
\bibitem [{\citenamefont {Delarue}\ \emph {et~al.}(2017)\citenamefont
  {Delarue}, \citenamefont {Koehl},\ and\ \citenamefont
  {Orland}}]{Delarue2017}%
  \BibitemOpen
  \bibfield  {author} {\bibinfo {author} {\bibfnamefont {M.}~\bibnamefont
  {Delarue}}, \bibinfo {author} {\bibfnamefont {P.}~\bibnamefont {Koehl}},\
  and\ \bibinfo {author} {\bibfnamefont {H.}~\bibnamefont {Orland}},\
  }\bibfield  {title} {\bibinfo {title} {Ab initio sampling of transition paths
  by conditioned {Langevin} dynamics},\ }\href
  {https://doi.org/10.1063/1.4985651} {\bibfield  {journal} {\bibinfo
  {journal} {The Journal of Chemical Physics}\ }\textbf {\bibinfo {volume}
  {147}},\ \bibinfo {pages} {152703} (\bibinfo {year} {2017})}\BibitemShut
  {NoStop}%
\bibitem [{\citenamefont {Elber}\ \emph {et~al.}(2020)\citenamefont {Elber},
  \citenamefont {Makarov},\ and\ \citenamefont {Orland}}]{Elber2020}%
  \BibitemOpen
  \bibfield  {author} {\bibinfo {author} {\bibfnamefont {R.}~\bibnamefont
  {Elber}}, \bibinfo {author} {\bibfnamefont {D.~E.}\ \bibnamefont {Makarov}},\
  and\ \bibinfo {author} {\bibfnamefont {H.}~\bibnamefont {Orland}},\
  }\href@noop {} {\emph {\bibinfo {title} {Molecular kinetics in condensed
  phases: Theory, simulation, and analysis}}}\ (\bibinfo  {publisher} {John
  Wiley \& Sons},\ \bibinfo {year} {2020})\BibitemShut {NoStop}%
\bibitem [{\citenamefont {Koehl}\ and\ \citenamefont
  {Orland}(2022)}]{Koehl2022}%
  \BibitemOpen
  \bibfield  {author} {\bibinfo {author} {\bibfnamefont {P.}~\bibnamefont
  {Koehl}}\ and\ \bibinfo {author} {\bibfnamefont {H.}~\bibnamefont {Orland}},\
  }\bibfield  {title} {\bibinfo {title} {Sampling constrained stochastic
  trajectories using {Brownian} bridges},\ }\href
  {https://doi.org/10.1063/5.0102295} {\bibfield  {journal} {\bibinfo
  {journal} {The Journal of Chemical Physics}\ }\textbf {\bibinfo {volume}
  {157}},\ \bibinfo {pages} {054105} (\bibinfo {year} {2022})}\BibitemShut
  {NoStop}%
\bibitem [{\citenamefont {Neupane}\ \emph {et~al.}(2012)\citenamefont
  {Neupane}, \citenamefont {Ritchie}, \citenamefont {Yu}, \citenamefont
  {Foster}, \citenamefont {Wang},\ and\ \citenamefont
  {Woodside}}]{Neupane2012}%
  \BibitemOpen
  \bibfield  {author} {\bibinfo {author} {\bibfnamefont {K.}~\bibnamefont
  {Neupane}}, \bibinfo {author} {\bibfnamefont {D.~B.}\ \bibnamefont
  {Ritchie}}, \bibinfo {author} {\bibfnamefont {H.}~\bibnamefont {Yu}},
  \bibinfo {author} {\bibfnamefont {D.~A.~N.}\ \bibnamefont {Foster}}, \bibinfo
  {author} {\bibfnamefont {F.}~\bibnamefont {Wang}},\ and\ \bibinfo {author}
  {\bibfnamefont {M.~T.}\ \bibnamefont {Woodside}},\ }\bibfield  {title}
  {\bibinfo {title} {Transition {Path} {Times} for {Nucleic} {Acid} {Folding}
  {Determined} from {Energy}-{Landscape} {Analysis} of {Single}-{Molecule}
  {Trajectories}},\ }\href {https://doi.org/10.1103/PhysRevLett.109.068102}
  {\bibfield  {journal} {\bibinfo  {journal} {Physical Review Letters}\
  }\textbf {\bibinfo {volume} {109}},\ \bibinfo {pages} {068102} (\bibinfo
  {year} {2012})}\BibitemShut {NoStop}%
\bibitem [{\citenamefont {Chung}\ and\ \citenamefont
  {Eaton}(2018)}]{Chung2018}%
  \BibitemOpen
  \bibfield  {author} {\bibinfo {author} {\bibfnamefont {H.~S.}\ \bibnamefont
  {Chung}}\ and\ \bibinfo {author} {\bibfnamefont {W.~A.}\ \bibnamefont
  {Eaton}},\ }\bibfield  {title} {\bibinfo {title} {Protein folding transition
  path times from single molecule {FRET}},\ }\href
  {https://doi.org/10.1016/j.sbi.2017.10.007} {\bibfield  {journal} {\bibinfo
  {journal} {Current Opinion in Structural Biology}\ }\textbf {\bibinfo
  {volume} {48}},\ \bibinfo {pages} {30} (\bibinfo {year} {2018})}\BibitemShut
  {NoStop}%
\bibitem [{\citenamefont {Faccioli}\ \emph {et~al.}(2006)\citenamefont
  {Faccioli}, \citenamefont {Sega}, \citenamefont {Pederiva},\ and\
  \citenamefont {Orland}}]{Faccioli2006}%
  \BibitemOpen
  \bibfield  {author} {\bibinfo {author} {\bibfnamefont {P.}~\bibnamefont
  {Faccioli}}, \bibinfo {author} {\bibfnamefont {M.}~\bibnamefont {Sega}},
  \bibinfo {author} {\bibfnamefont {F.}~\bibnamefont {Pederiva}},\ and\
  \bibinfo {author} {\bibfnamefont {H.}~\bibnamefont {Orland}},\ }\bibfield
  {title} {\bibinfo {title} {Dominant {Pathways} in {Protein} {Folding}},\
  }\href {https://doi.org/10.1103/PhysRevLett.97.108101} {\bibfield  {journal}
  {\bibinfo  {journal} {Physical Review Letters}\ }\textbf {\bibinfo {volume}
  {97}},\ \bibinfo {pages} {108101} (\bibinfo {year} {2006})}\BibitemShut
  {NoStop}%
\bibitem [{\citenamefont {Autieri}\ \emph {et~al.}(2009)\citenamefont
  {Autieri}, \citenamefont {Faccioli}, \citenamefont {Sega}, \citenamefont
  {Pederiva},\ and\ \citenamefont {Orland}}]{Autieri2009}%
  \BibitemOpen
  \bibfield  {author} {\bibinfo {author} {\bibfnamefont {E.}~\bibnamefont
  {Autieri}}, \bibinfo {author} {\bibfnamefont {P.}~\bibnamefont {Faccioli}},
  \bibinfo {author} {\bibfnamefont {M.}~\bibnamefont {Sega}}, \bibinfo {author}
  {\bibfnamefont {F.}~\bibnamefont {Pederiva}},\ and\ \bibinfo {author}
  {\bibfnamefont {H.}~\bibnamefont {Orland}},\ }\bibfield  {title} {\bibinfo
  {title} {Dominant reaction pathways in high-dimensional systems},\ }\href
  {https://doi.org/10.1063/1.3074271} {\bibfield  {journal} {\bibinfo
  {journal} {The Journal of Chemical Physics}\ }\textbf {\bibinfo {volume}
  {130}},\ \bibinfo {pages} {064106} (\bibinfo {year} {2009})}\BibitemShut
  {NoStop}%
\bibitem [{\citenamefont {Faccioli}\ \emph {et~al.}(2010)\citenamefont
  {Faccioli}, \citenamefont {Lonardi},\ and\ \citenamefont
  {Orland}}]{Faccioli2010}%
  \BibitemOpen
  \bibfield  {author} {\bibinfo {author} {\bibfnamefont {P.}~\bibnamefont
  {Faccioli}}, \bibinfo {author} {\bibfnamefont {A.}~\bibnamefont {Lonardi}},\
  and\ \bibinfo {author} {\bibfnamefont {H.}~\bibnamefont {Orland}},\
  }\bibfield  {title} {\bibinfo {title} {Dominant reaction pathways in protein
  folding: {A} direct validation against molecular dynamics simulations},\
  }\href {https://doi.org/10.1063/1.3459097} {\bibfield  {journal} {\bibinfo
  {journal} {The Journal of Chemical Physics}\ }\textbf {\bibinfo {volume}
  {133}},\ \bibinfo {pages} {045104} (\bibinfo {year} {2010})}\BibitemShut
  {NoStop}%
\bibitem [{\citenamefont {McDole}\ \emph {et~al.}(2018)\citenamefont {McDole},
  \citenamefont {Guignard}, \citenamefont {Amat}, \citenamefont {Berger},
  \citenamefont {Malandain}, \citenamefont {Royer}, \citenamefont {Turaga},
  \citenamefont {Branson},\ and\ \citenamefont {Keller}}]{McDole2018}%
  \BibitemOpen
  \bibfield  {author} {\bibinfo {author} {\bibfnamefont {K.}~\bibnamefont
  {McDole}}, \bibinfo {author} {\bibfnamefont {L.}~\bibnamefont {Guignard}},
  \bibinfo {author} {\bibfnamefont {F.}~\bibnamefont {Amat}}, \bibinfo {author}
  {\bibfnamefont {A.}~\bibnamefont {Berger}}, \bibinfo {author} {\bibfnamefont
  {G.}~\bibnamefont {Malandain}}, \bibinfo {author} {\bibfnamefont {L.~A.}\
  \bibnamefont {Royer}}, \bibinfo {author} {\bibfnamefont {S.~C.}\ \bibnamefont
  {Turaga}}, \bibinfo {author} {\bibfnamefont {K.}~\bibnamefont {Branson}},\
  and\ \bibinfo {author} {\bibfnamefont {P.~J.}\ \bibnamefont {Keller}},\
  }\bibfield  {title} {\bibinfo {title} {In {Toto} {Imaging} and
  {Reconstruction} of {Post}-{Implantation} {Mouse} {Development} at the
  {Single}-{Cell} {Level}},\ }\href
  {https://doi.org/10.1016/j.cell.2018.09.031} {\bibfield  {journal} {\bibinfo
  {journal} {Cell}\ }\textbf {\bibinfo {volume} {175}},\ \bibinfo {pages} {859}
  (\bibinfo {year} {2018})}\BibitemShut {NoStop}%
\bibitem [{\citenamefont {Lange}\ \emph {et~al.}(2023)\citenamefont {Lange}
  \emph {et~al.}}]{Lange2023}%
  \BibitemOpen
  \bibfield  {author} {\bibinfo {author} {\bibfnamefont {M.}~\bibnamefont
  {Lange}} \emph {et~al.},\ }\bibfield  {title} {\bibinfo {title} {Zebrahub –
  {Multimodal} {Zebrafish} {Developmental} {Atlas} {Reveals} the
  {State}-{Transition} {Dynamics} of {Late}-{Vertebrate} {Pluripotent} {Axial}
  {Progenitors}},\ }\href {https://doi.org/10.1101/2023.03.06.531398}
  {\bibfield  {journal} {\bibinfo  {journal} {bioRxiv}\ ,\ \bibinfo {pages}
  {2023.03.06.531398}} (\bibinfo {year} {2023})}\BibitemShut {NoStop}%
\bibitem [{\citenamefont {Sandberg}(2014)}]{Sandberg2014}%
  \BibitemOpen
  \bibfield  {author} {\bibinfo {author} {\bibfnamefont {R.}~\bibnamefont
  {Sandberg}},\ }\bibfield  {title} {\bibinfo {title} {Entering the era of
  single-cell transcriptomics in biology and medicine},\ }\href
  {https://doi.org/10.1038/nmeth.2764} {\bibfield  {journal} {\bibinfo
  {journal} {Nature Methods}\ }\textbf {\bibinfo {volume} {11}},\ \bibinfo
  {pages} {22} (\bibinfo {year} {2014})}\BibitemShut {NoStop}%
\bibitem [{\citenamefont {Gawad}\ \emph {et~al.}(2016)\citenamefont {Gawad},
  \citenamefont {Koh},\ and\ \citenamefont {Quake}}]{Gawad2016}%
  \BibitemOpen
  \bibfield  {author} {\bibinfo {author} {\bibfnamefont {C.}~\bibnamefont
  {Gawad}}, \bibinfo {author} {\bibfnamefont {W.}~\bibnamefont {Koh}},\ and\
  \bibinfo {author} {\bibfnamefont {S.~R.}\ \bibnamefont {Quake}},\ }\bibfield
  {title} {\bibinfo {title} {Single-cell genome sequencing: current state of
  the science},\ }\href {https://doi.org/10.1038/nrg.2015.16} {\bibfield
  {journal} {\bibinfo  {journal} {Nature Reviews Genetics}\ }\textbf {\bibinfo
  {volume} {17}},\ \bibinfo {pages} {175} (\bibinfo {year} {2016})}\BibitemShut
  {NoStop}%
\bibitem [{\citenamefont {Schaum}\ \emph {et~al.}(2018)\citenamefont {Schaum}
  \emph {et~al.}}]{Schaum2018}%
  \BibitemOpen
  \bibfield  {author} {\bibinfo {author} {\bibfnamefont {N.}~\bibnamefont
  {Schaum}} \emph {et~al.},\ }\bibfield  {title} {\bibinfo {title} {Single-cell
  transcriptomics of 20 mouse organs creates a \textit{{Tabula} {Muris}}},\
  }\href {https://doi.org/10.1038/s41586-018-0590-4} {\bibfield  {journal}
  {\bibinfo  {journal} {Nature}\ }\textbf {\bibinfo {volume} {562}},\ \bibinfo
  {pages} {367} (\bibinfo {year} {2018})}\BibitemShut {NoStop}%
\bibitem [{\citenamefont {Jones}\ \emph {et~al.}(2022)\citenamefont {Jones}
  \emph {et~al.}}]{Jones2022}%
  \BibitemOpen
  \bibfield  {author} {\bibinfo {author} {\bibfnamefont {R.~C.}\ \bibnamefont
  {Jones}} \emph {et~al.},\ }\bibfield  {title} {\bibinfo {title} {The
  \textit{{Tabula} {Sapiens}}: {A} multiple-organ, single-cell transcriptomic
  atlas of humans},\ }\bibfield  {journal} {\bibinfo  {journal} {Science}\
  }\href {https://doi.org/10.1126/science.abl4896} {10.1126/science.abl4896}
  (\bibinfo {year} {2022})\BibitemShut {NoStop}%
\bibitem [{\citenamefont {La~Manno}\ \emph {et~al.}(2018)\citenamefont
  {La~Manno}, \citenamefont {Soldatov}, \citenamefont {Zeisel}, \citenamefont
  {Braun}, \citenamefont {Hochgerner}, \citenamefont {Petukhov}, \citenamefont
  {Lidschreiber}, \citenamefont {Kastriti}, \citenamefont {L\"{o}nnerberg},
  \citenamefont {Furlan}, \citenamefont {Fan}, \citenamefont {Borm},
  \citenamefont {Liu}, \citenamefont {van Bruggen}, \citenamefont {Guo},
  \citenamefont {He}, \citenamefont {Barker}, \citenamefont {Sundstr\"{o}m},
  \citenamefont {Castelo-Branco}, \citenamefont {Cramer}, \citenamefont
  {Adameyko}, \citenamefont {Linnarsson},\ and\ \citenamefont
  {Kharchenko}}]{LaManno2018}%
  \BibitemOpen
  \bibfield  {author} {\bibinfo {author} {\bibfnamefont {G.}~\bibnamefont
  {La~Manno}}, \bibinfo {author} {\bibfnamefont {R.}~\bibnamefont {Soldatov}},
  \bibinfo {author} {\bibfnamefont {A.}~\bibnamefont {Zeisel}}, \bibinfo
  {author} {\bibfnamefont {E.}~\bibnamefont {Braun}}, \bibinfo {author}
  {\bibfnamefont {H.}~\bibnamefont {Hochgerner}}, \bibinfo {author}
  {\bibfnamefont {V.}~\bibnamefont {Petukhov}}, \bibinfo {author}
  {\bibfnamefont {K.}~\bibnamefont {Lidschreiber}}, \bibinfo {author}
  {\bibfnamefont {M.~E.}\ \bibnamefont {Kastriti}}, \bibinfo {author}
  {\bibfnamefont {P.}~\bibnamefont {L\"{o}nnerberg}}, \bibinfo {author}
  {\bibfnamefont {A.}~\bibnamefont {Furlan}}, \bibinfo {author} {\bibfnamefont
  {J.}~\bibnamefont {Fan}}, \bibinfo {author} {\bibfnamefont {L.~E.}\
  \bibnamefont {Borm}}, \bibinfo {author} {\bibfnamefont {Z.}~\bibnamefont
  {Liu}}, \bibinfo {author} {\bibfnamefont {D.}~\bibnamefont {van Bruggen}},
  \bibinfo {author} {\bibfnamefont {J.}~\bibnamefont {Guo}}, \bibinfo {author}
  {\bibfnamefont {X.}~\bibnamefont {He}}, \bibinfo {author} {\bibfnamefont
  {R.}~\bibnamefont {Barker}}, \bibinfo {author} {\bibfnamefont
  {E.}~\bibnamefont {Sundstr\"{o}m}}, \bibinfo {author} {\bibfnamefont
  {G.}~\bibnamefont {Castelo-Branco}}, \bibinfo {author} {\bibfnamefont
  {P.}~\bibnamefont {Cramer}}, \bibinfo {author} {\bibfnamefont
  {I.}~\bibnamefont {Adameyko}}, \bibinfo {author} {\bibfnamefont
  {S.}~\bibnamefont {Linnarsson}},\ and\ \bibinfo {author} {\bibfnamefont
  {P.~V.}\ \bibnamefont {Kharchenko}},\ }\bibfield  {title} {\bibinfo {title}
  {{RNA} velocity of single cells},\ }\href
  {https://doi.org/10.1038/s41586-018-0414-6} {\bibfield  {journal} {\bibinfo
  {journal} {Nature}\ }\textbf {\bibinfo {volume} {560}},\ \bibinfo {pages}
  {494} (\bibinfo {year} {2018})}\BibitemShut {NoStop}%
\bibitem [{\citenamefont {Bergen}\ \emph {et~al.}(2020)\citenamefont {Bergen},
  \citenamefont {Lange}, \citenamefont {Peidli}, \citenamefont {Wolf},\ and\
  \citenamefont {Theis}}]{Bergen2020}%
  \BibitemOpen
  \bibfield  {author} {\bibinfo {author} {\bibfnamefont {V.}~\bibnamefont
  {Bergen}}, \bibinfo {author} {\bibfnamefont {M.}~\bibnamefont {Lange}},
  \bibinfo {author} {\bibfnamefont {S.}~\bibnamefont {Peidli}}, \bibinfo
  {author} {\bibfnamefont {F.~A.}\ \bibnamefont {Wolf}},\ and\ \bibinfo
  {author} {\bibfnamefont {F.~J.}\ \bibnamefont {Theis}},\ }\bibfield  {title}
  {\bibinfo {title} {Generalizing {RNA} velocity to transient cell states
  through dynamical modeling},\ }\href
  {https://doi.org/10.1038/s41587-020-0591-3} {\bibfield  {journal} {\bibinfo
  {journal} {Nature Biotechnology}\ }\textbf {\bibinfo {volume} {38}},\
  \bibinfo {pages} {1408} (\bibinfo {year} {2020})}\BibitemShut {NoStop}%
\bibitem [{\citenamefont {Lange}\ \emph {et~al.}(2022)\citenamefont {Lange},
  \citenamefont {Bergen}, \citenamefont {Klein}, \citenamefont {Setty},
  \citenamefont {Reuter}, \citenamefont {Bakhti}, \citenamefont {Lickert},
  \citenamefont {Ansari}, \citenamefont {Schniering}, \citenamefont {Schiller},
  \citenamefont {Pe'er},\ and\ \citenamefont {Theis}}]{Lange2022}%
  \BibitemOpen
  \bibfield  {author} {\bibinfo {author} {\bibfnamefont {M.}~\bibnamefont
  {Lange}}, \bibinfo {author} {\bibfnamefont {V.}~\bibnamefont {Bergen}},
  \bibinfo {author} {\bibfnamefont {M.}~\bibnamefont {Klein}}, \bibinfo
  {author} {\bibfnamefont {M.}~\bibnamefont {Setty}}, \bibinfo {author}
  {\bibfnamefont {B.}~\bibnamefont {Reuter}}, \bibinfo {author} {\bibfnamefont
  {M.}~\bibnamefont {Bakhti}}, \bibinfo {author} {\bibfnamefont
  {H.}~\bibnamefont {Lickert}}, \bibinfo {author} {\bibfnamefont
  {M.}~\bibnamefont {Ansari}}, \bibinfo {author} {\bibfnamefont
  {J.}~\bibnamefont {Schniering}}, \bibinfo {author} {\bibfnamefont {H.~B.}\
  \bibnamefont {Schiller}}, \bibinfo {author} {\bibfnamefont {D.}~\bibnamefont
  {Pe'er}},\ and\ \bibinfo {author} {\bibfnamefont {F.~J.}\ \bibnamefont
  {Theis}},\ }\bibfield  {title} {\bibinfo {title} {{CellRank} for directed
  single-cell fate mapping},\ }\href
  {https://doi.org/10.1038/s41592-021-01346-6} {\bibfield  {journal} {\bibinfo
  {journal} {Nature Methods}\ }\textbf {\bibinfo {volume} {19}},\ \bibinfo
  {pages} {159} (\bibinfo {year} {2022})}\BibitemShut {NoStop}%
\bibitem [{\citenamefont {Weiler}\ \emph {et~al.}(2023)\citenamefont {Weiler},
  \citenamefont {Lange}, \citenamefont {Klein}, \citenamefont {Pe'er},\ and\
  \citenamefont {Theis}}]{Weiler2023}%
  \BibitemOpen
  \bibfield  {author} {\bibinfo {author} {\bibfnamefont {P.}~\bibnamefont
  {Weiler}}, \bibinfo {author} {\bibfnamefont {M.}~\bibnamefont {Lange}},
  \bibinfo {author} {\bibfnamefont {M.}~\bibnamefont {Klein}}, \bibinfo
  {author} {\bibfnamefont {D.}~\bibnamefont {Pe'er}},\ and\ \bibinfo {author}
  {\bibfnamefont {F.}~\bibnamefont {Theis}},\ }\bibfield  {title} {\bibinfo
  {title} {Unified fate mapping in multiview single-cell data},\ }\href@noop {}
  {\bibfield  {journal} {\bibinfo  {journal} {bioRxiv}\ ,\ \bibinfo {pages}
  {2023}} (\bibinfo {year} {2023})}\BibitemShut {NoStop}%
\bibitem [{\citenamefont {Van~Kampen}(1992)}]{1992vanKampen_stochastic}%
  \BibitemOpen
  \bibfield  {author} {\bibinfo {author} {\bibfnamefont {N.~G.}\ \bibnamefont
  {Van~Kampen}},\ }\href@noop {} {\emph {\bibinfo {title} {Stochastic processes
  in physics and chemistry}}},\ Vol.~\bibinfo {volume} {1}\ (\bibinfo
  {publisher} {Elsevier},\ \bibinfo {year} {1992})\BibitemShut {NoStop}%
\bibitem [{\citenamefont {Gillespie}(1976)}]{Gillespie1976}%
  \BibitemOpen
  \bibfield  {author} {\bibinfo {author} {\bibfnamefont {D.~T.}\ \bibnamefont
  {Gillespie}},\ }\bibfield  {title} {\bibinfo {title} {A general method for
  numerically simulating the stochastic time evolution of coupled chemical
  reactions},\ }\href {https://doi.org/10.1016/0021-9991(76)90041-3} {\bibfield
   {journal} {\bibinfo  {journal} {Journal of Computational Physics}\ }\textbf
  {\bibinfo {volume} {22}},\ \bibinfo {pages} {403} (\bibinfo {year}
  {1976})}\BibitemShut {NoStop}%
\bibitem [{\citenamefont {M\"{u}ller}\ and\ \citenamefont
  {Brown}(1979)}]{Mueller1979}%
  \BibitemOpen
  \bibfield  {author} {\bibinfo {author} {\bibfnamefont {K.}~\bibnamefont
  {M\"{u}ller}}\ and\ \bibinfo {author} {\bibfnamefont {L.~D.}\ \bibnamefont
  {Brown}},\ }\bibfield  {title} {\bibinfo {title} {Location of saddle points
  and minimum energy paths by a constrained simplex optimization procedure},\
  }\href {https://doi.org/10.1007/BF00547608} {\bibfield  {journal} {\bibinfo
  {journal} {Theoretica chimica acta}\ }\textbf {\bibinfo {volume} {53}},\
  \bibinfo {pages} {75} (\bibinfo {year} {1979})}\BibitemShut {NoStop}%
\bibitem [{\citenamefont {M\"{u}ller}(1980)}]{Mueller1980}%
  \BibitemOpen
  \bibfield  {author} {\bibinfo {author} {\bibfnamefont {K.}~\bibnamefont
  {M\"{u}ller}},\ }\bibfield  {title} {\bibinfo {title} {Reaction {Paths} on
  {Multidimensional} {Energy} {Hypersurfaces}},\ }\href
  {https://doi.org/10.1002/anie.198000013} {\bibfield  {journal} {\bibinfo
  {journal} {Angewandte Chemie International Edition in English}\ }\textbf
  {\bibinfo {volume} {19}},\ \bibinfo {pages} {1} (\bibinfo {year}
  {1980})}\BibitemShut {NoStop}%
\bibitem [{\citenamefont {Binder}\ \emph {et~al.}(1992)\citenamefont {Binder},
  \citenamefont {Heermann},\ and\ \citenamefont {Binder}}]{Binder1992}%
  \BibitemOpen
  \bibfield  {author} {\bibinfo {author} {\bibfnamefont {K.}~\bibnamefont
  {Binder}}, \bibinfo {author} {\bibfnamefont {D.~W.}\ \bibnamefont
  {Heermann}},\ and\ \bibinfo {author} {\bibfnamefont {K.}~\bibnamefont
  {Binder}},\ }\href@noop {} {\emph {\bibinfo {title} {Monte Carlo simulation
  in statistical physics}}},\ Vol.~\bibinfo {volume} {8}\ (\bibinfo
  {publisher} {Springer},\ \bibinfo {year} {1992})\BibitemShut {NoStop}%
\bibitem [{\citenamefont {Bastidas-Ponce}\ \emph {et~al.}(2019)\citenamefont
  {Bastidas-Ponce}, \citenamefont {Tritschler}, \citenamefont {Dony},
  \citenamefont {Scheibner}, \citenamefont {Tarquis-Medina}, \citenamefont
  {Salinno}, \citenamefont {Schirge}, \citenamefont {Burtscher}, \citenamefont
  {Böttcher}, \citenamefont {Theis}, \citenamefont {Lickert},\ and\
  \citenamefont {Bakhti}}]{BastidasPonce2019}%
  \BibitemOpen
  \bibfield  {author} {\bibinfo {author} {\bibfnamefont {A.}~\bibnamefont
  {Bastidas-Ponce}}, \bibinfo {author} {\bibfnamefont {S.}~\bibnamefont
  {Tritschler}}, \bibinfo {author} {\bibfnamefont {L.}~\bibnamefont {Dony}},
  \bibinfo {author} {\bibfnamefont {K.}~\bibnamefont {Scheibner}}, \bibinfo
  {author} {\bibfnamefont {M.}~\bibnamefont {Tarquis-Medina}}, \bibinfo
  {author} {\bibfnamefont {C.}~\bibnamefont {Salinno}}, \bibinfo {author}
  {\bibfnamefont {S.}~\bibnamefont {Schirge}}, \bibinfo {author} {\bibfnamefont
  {I.}~\bibnamefont {Burtscher}}, \bibinfo {author} {\bibfnamefont
  {A.}~\bibnamefont {Böttcher}}, \bibinfo {author} {\bibfnamefont {F.~J.}\
  \bibnamefont {Theis}}, \bibinfo {author} {\bibfnamefont {H.}~\bibnamefont
  {Lickert}},\ and\ \bibinfo {author} {\bibfnamefont {M.}~\bibnamefont
  {Bakhti}},\ }\bibfield  {title} {\bibinfo {title} {Comprehensive single cell
  {mRNA} profiling reveals a detailed roadmap for pancreatic
  endocrinogenesis},\ }\href {https://doi.org/10.1242/dev.173849} {\bibfield
  {journal} {\bibinfo  {journal} {Development}\ }\textbf {\bibinfo {volume}
  {146}},\ \bibinfo {pages} {dev173849} (\bibinfo {year} {2019})}\BibitemShut
  {NoStop}%
\bibitem [{\citenamefont {McInnes}\ \emph {et~al.}(2018)\citenamefont
  {McInnes}, \citenamefont {Healy},\ and\ \citenamefont
  {Melville}}]{McInnes2018}%
  \BibitemOpen
  \bibfield  {author} {\bibinfo {author} {\bibfnamefont {L.}~\bibnamefont
  {McInnes}}, \bibinfo {author} {\bibfnamefont {J.}~\bibnamefont {Healy}},\
  and\ \bibinfo {author} {\bibfnamefont {J.}~\bibnamefont {Melville}},\
  }\bibfield  {title} {\bibinfo {title} {{UMAP}: {Uniform} {Manifold}
  {Approximation} and {Projection} for {Dimension} {Reduction}},\ }\href
  {https://doi.org/10.48550/arXiv.1802.03426} {\bibfield  {journal} {\bibinfo
  {journal} {arXiv}\ ,\ \bibinfo {pages} {1802.0342}} (\bibinfo {year}
  {2018})}\BibitemShut {NoStop}%
\bibitem [{\citenamefont {Becht}\ \emph {et~al.}(2019)\citenamefont {Becht},
  \citenamefont {McInnes}, \citenamefont {Healy}, \citenamefont {Dutertre},
  \citenamefont {Kwok}, \citenamefont {Ng}, \citenamefont {Ginhoux},\ and\
  \citenamefont {Newell}}]{Becht2019}%
  \BibitemOpen
  \bibfield  {author} {\bibinfo {author} {\bibfnamefont {E.}~\bibnamefont
  {Becht}}, \bibinfo {author} {\bibfnamefont {L.}~\bibnamefont {McInnes}},
  \bibinfo {author} {\bibfnamefont {J.}~\bibnamefont {Healy}}, \bibinfo
  {author} {\bibfnamefont {C.-A.}\ \bibnamefont {Dutertre}}, \bibinfo {author}
  {\bibfnamefont {I.~W.~H.}\ \bibnamefont {Kwok}}, \bibinfo {author}
  {\bibfnamefont {L.~G.}\ \bibnamefont {Ng}}, \bibinfo {author} {\bibfnamefont
  {F.}~\bibnamefont {Ginhoux}},\ and\ \bibinfo {author} {\bibfnamefont {E.~W.}\
  \bibnamefont {Newell}},\ }\bibfield  {title} {\bibinfo {title}
  {Dimensionality reduction for visualizing single-cell data using {UMAP}},\
  }\href {https://doi.org/10.1038/nbt.4314} {\bibfield  {journal} {\bibinfo
  {journal} {Nature Biotechnology}\ }\textbf {\bibinfo {volume} {37}},\
  \bibinfo {pages} {38} (\bibinfo {year} {2019})}\BibitemShut {NoStop}%
\bibitem [{\citenamefont {Moler}\ and\ \citenamefont
  {Van~Loan}(2003)}]{Moler2003}%
  \BibitemOpen
  \bibfield  {author} {\bibinfo {author} {\bibfnamefont {C.}~\bibnamefont
  {Moler}}\ and\ \bibinfo {author} {\bibfnamefont {C.}~\bibnamefont
  {Van~Loan}},\ }\bibfield  {title} {\bibinfo {title} {Nineteen dubious ways to
  compute the exponential of a matrix, twenty-five years later},\ }\href@noop
  {} {\bibfield  {journal} {\bibinfo  {journal} {SIAM review}\ }\textbf
  {\bibinfo {volume} {45}},\ \bibinfo {pages} {3} (\bibinfo {year}
  {2003})}\BibitemShut {NoStop}%
\bibitem [{\citenamefont {Golub}\ and\ \citenamefont
  {Van~Loan}(2013)}]{Golub2013}%
  \BibitemOpen
  \bibfield  {author} {\bibinfo {author} {\bibfnamefont {G.~H.}\ \bibnamefont
  {Golub}}\ and\ \bibinfo {author} {\bibfnamefont {C.~F.}\ \bibnamefont
  {Van~Loan}},\ }\href@noop {} {\emph {\bibinfo {title} {Matrix
  computations}}}\ (\bibinfo  {publisher} {JHU press},\ \bibinfo {year}
  {2013})\BibitemShut {NoStop}%
\end{thebibliography}%

\clearpage
\appendix
\small
\section{Pseudo code implementations}
\label{sec:pseudocode}

\Cref{al:gillespie} is the implementation we have used in this work to
sample Markov bridges. However, as mentioned in the main text, for large systems it might not be adequate due to the computational cost of the eigenvalue decomposition of the $\W$ matrix. We therefore give~\cref{al:markov_chain}, which is an
implementation not relying on a matrix decomposition of the form $\ee^{t \W} = U
\ee^{-t D} U^{-1}$ to evaluate $Q(t)$. In this approach, $Q(t)$ can be evaluated
at specified discrete times $t_j = j \Delta t$ (with $t_M = t_\text{f}$) by
integrating~\cref{eq:master_equation_backward}. We note here that
matrix-vector multiplications can be performed very quickly with modern
GPUs, which can be leveraged in an integration scheme. The time resolution $\Delta t$ should be chosen so that $\Delta t <
\Delta t^\ast = 1 / \max_\alpha{\left( \Gamma_\alpha \right)}$
($\Gamma_\alpha$ is the escape rate as defined in~\cref{eq:escape_rate}).

\begin{algorithm}[H]
  \caption{Gillespie sampling for Markov bridges}
  \label{al:gillespie}
  \begin{algorithmic}
    \REQUIRE $(\alpha_\text{i}, 0)$, rate matrix $W$
    \STATE Set $(\alpha, t) \leftarrow (\alpha_\text{i}, 0)$
    \PRINT $(\alpha, t)$.
    \WHILE{$t < t_\text{f}$}
      \STATE Draw $u_1$ and $u_2$ as $\mathcal{U}(0,1)$.
      \STATE $\tau \leftarrow$ Solve~\cref{eq:dwell_time_draw} with $u_1$.
      $Q(t)$ is computed through~\cref{eq:qeval}.
      \STATE Take $\gamma$ so that $\sum \limits_{\beta=1}^{\gamma-1}
      V_{\beta \alpha}<
      u_2 \Gamma_\alpha(t+\tau) \le \sum \limits_{\beta=1}^{\gamma}
      V_{\beta \alpha}$.
      \STATE Update $(\alpha, t) \leftarrow (\gamma, t+\tau)$.
      \PRINT $(\alpha, t)$.
    \ENDWHILE
  \end{algorithmic}
\end{algorithm}

\begin{algorithm}[H]
  \caption{Standard sampling for Markov bridges}
  \label{al:markov_chain}
  \begin{algorithmic}
    \REQUIRE $(\alpha_\text{i}, 0)$, rate matrix $W$, time resolution $\Delta t = t_\text{f} / M$
    \STATE Integrate~\cref{eq:master_equation_backward} and evaluate $Q(t_j),\, \forall j \in \llbracket 0, M \rrbracket$.
    \STATE Set $(\alpha, t) \leftarrow (\alpha_\text{i}, 0)$
    \PRINT $(\alpha, t)$.
    \WHILE{$t < t_\text{f}$}
      \STATE Draw $u_1$ and $u_2$ as $\mathcal{U}(0,1)$.
      \STATE $\tau \leftarrow j \Delta t$, where $j$ is the smallest integer such that~\cref{eq:dwell_time_draw} is positive (with $u_1$).
      \STATE Take $\gamma$ so that $\sum \limits_{\beta=1}^{\gamma-1} V_{\beta \alpha}< u_2 \Gamma_\alpha(t+\tau) \le \sum \limits_{\beta=1}^{\gamma} V_{\beta \alpha}$.
      \STATE Update $(\alpha, t) \leftarrow (\gamma, t+\tau)$.
      \PRINT $(\alpha, t)$.
    \ENDWHILE
  \end{algorithmic}
\end{algorithm}

\section{Analytical treatment of diffusion on the 1D lattice}
\label{sec:diffusion}
To solve~\cref{eq:diffusion_mequation_P}, we introduce the discrete
Fourier transform $\tilde{P}_k$, such that:
\begin{align}
  \begin{aligned}
    \tilde{P}_k &= \sum \limits_{n=1}^{N} P_n \ee^{-\ii \frac{2 \pi}{N} n k}, \\
    P_n &= \frac{1}{N} \sum \limits_{k=1}^{N} \tilde{P}_k \ee^{\ii \frac{2
    \pi}{N} n k}.
  \end{aligned}
  \label{eq:diffusion_dft_p}
\end{align}

Taking the time derivative of $\tilde{P}_k$, and
using~\cref{eq:diffusion_mequation_P}, we obtain the following
system of ODE:
\begin{align}
  \begin{aligned}
    \frac{\dd \tilde{P}_k}{\dd t} &= - \tilde{\omega}_k \tilde{P}_k, \qquad \forall k \in \llbracket 1, N \rrbracket,
  \end{aligned}
  \label{eq:diffusion_mequation_Ptilde}
\end{align}
where we introduced:
\begin{align}
  \begin{aligned}
    \tilde{\omega}_k &= - 2 \ii \sin{\left( \frac{\pi k}{N} \right)} \left(
    \beta \ee^{\ii \frac{\pi k}{N}} - \alpha \ee^{-\ii \frac{\pi k}{N}}
    \right), \\
    \omega_n &= (\alpha + \beta) \delta_n - \alpha \delta_{n-1} - \beta
    \delta_{n+1},
  \end{aligned}
\end{align}
where $\delta_n$ is the Kronecker delta. In general, $\tilde{\omega}_k$
has a non-zero imaginary part. However,
when $\alpha = \beta$, it reduces to $\tilde{\omega}_k~=~4 \alpha
\sin^2{\left( \frac{\pi k}{N} \right)}$.

\Cref{eq:diffusion_mequation_Ptilde} is straightforwardly integrated and
yields:
\begin{align}
  \tilde{P}_k(t) = \ee^{-\tilde{\omega}_k t} \ee^{-\ii \frac{2 \pi}{N} n_\text{i} k},
  \label{eq:diffusion_ptilde}
\end{align}
where we used the initial condition $P_n(0) = \delta_{n,n_\text{i}}$.

For the backward master equation, we introduce $Q_n(t) =
\mathbbm{P}(n_\text{f}, t_\text{f} | n, t)$. \Cref{eq:master_equation_backward} reads:
\begin{align}
  \frac{\dd Q_n}{\dd t} = -\alpha Q_{n+1} - \beta Q_{n-1} + (\alpha +
  \beta) Q_n, \qquad \forall n \in \llbracket 1, N \rrbracket,
  \label{eq:diffusion_mequation_Q}
\end{align}
and its discrete Fourier transform satisfies the ODE:
\begin{align}
  \begin{aligned}
    \frac{\dd \tilde{Q}_k}{\dd t} &= \tilde{\omega}_k^\ast \tilde{Q}_k,
    \qquad \forall k \in \llbracket 1, N \rrbracket,
  \end{aligned}
  \label{eq:diffusion_mequation_Qtilde}
\end{align}
where the superscript $\ast$ denotes complex conjugation.
\Cref{eq:diffusion_mequation_Qtilde} is again straightforwardly integrated
and yields:
\begin{align}
  \tilde{Q}_k(t) = \ee^{-\tilde{\omega}_k^\ast (t_\text{f} - t)} \ee^{-\ii \frac{2 \pi}{N}
  n_\text{f} k}, \qquad \forall k \in \llbracket 1, N \rrbracket.
  \label{eq:diffusion_qtilde}
\end{align}
where we used the final condition $Q_n(t_\text{f}) = \delta_{n,n_\text{f}}$.

We can now compute the joint probability $R_n(t)~\propto~\mathbbm{P}(n_\text{f}, t_\text{f} |
n, t) \mathbbm{P}(n, t | n_\text{i}, 0)$:
\begin{align}
  \begin{aligned}
    R_n(t) &= \frac{1}{Z} P_n(t) Q_n(t), \qquad \forall n \in \llbracket 1, N
    \rrbracket, \\
    \tilde{R}_k &=
    \frac{\tilde{P}\ast\tilde{Q}_k}{\tilde{P}\ast\tilde{Q}_0}, \qquad
    \forall k \in \llbracket 1, N \rrbracket,
  \end{aligned}
  \label{eq:diffusion_r_rtilde}
\end{align}
where $Z = \sum_n P_n(t) Q_n(t) = P_{n_\text{f}}(t_f) = Q_{n_\text{i}}(0)$, and $\ast$
in this context is the convolution operator ($U\ast V_n = \sum_m U_{n-m} V_m$). Therefore
$R_n(t)$ can be exactly computed
from~\cref{eq:diffusion_ptilde,eq:diffusion_qtilde,eq:diffusion_r_rtilde}.

By differentiating $R_n(t)$ and
making use of the forward (\cref{eq:diffusion_mequation_P}) and backward
(\cref{eq:diffusion_mequation_Q}) master equations, we obtain the bridge
master equation:
\begin{align}
  \begin{aligned}
  \frac{\dd R_n}{ \dd t} &= \alpha \left(\frac{Q_{n}}{Q_{n-1}} R_{n-1} -
  \frac{Q_{n+1}}{Q_{n}} R_{n} \right)
  \\ &+ \beta \left(\frac{Q_{n}}{Q_{n+1}} R_{n+1} -
  \frac{Q_{n-1}}{Q_{n}} R_{n} \right), \qquad \forall n \in \llbracket 1,
  N \rrbracket.
  \end{aligned}
  \label{eq:diffusion_mequation_R}
\end{align}

The bridge master equation~\cref{eq:diffusion_mequation_R} has the same
form as~\cref{eq:master_equation_bridge}, where the forward and backward
rates are now time-dependent:
\begin{align}
  \begin{aligned}
    V_{n+1,n} = \alpha \frac{Q_{n+1}}{Q_{n}}, \qquad V_{n+p,n} = 0 \quad \text{
      if} \quad p > 1, \\
    V_{n-1,n} = \beta \frac{Q_{n-1}}{Q_{n}}, \qquad V_{n-p,n} = 0 \quad \text{
      if} \quad p > 1.
  \end{aligned}
  \label{eq:diffusion_bridge_rates}
\end{align}

\section{Mean and most likely trajectories}
\label{sec:mean_mostl_traj}

The mean trajectory shown in~\cref{fig:mb} was calculated by averaging
trajectories after time registration. First, we defined a suitable
subdivision of times $t_j = j \Delta t$. For a given trajectory $\lbrace(t_k,
\alpha_k)\rbrace$, the state at time $t_j$, namely $\alpha_j$ is such
that:
\begin{align}
  \alpha_j := \alpha_{k_j}, \quad \text{with} \quad t_{k_j} \le t_j < t_{k_j + 1}.
  \label{eq:state_registration}
\end{align}

The mean trajectory was obtained by averaging over the coordinates of the
$S$ trajectories:
\begin{align}
  \bar{\vec{x}}_j = \frac{1}{S} \sum \limits_{s=1}^{S} \vec{x}_{\alpha_j^{(s)}},
\end{align}
where $\vec{x}_\alpha$ is the coordinates vector associated with state $\alpha$
and $\alpha_j^{(s)}$ is the state of trajectory $s$ at time $t_j$ after
registration as defined in~\cref{eq:state_registration}.

The mean transition path shown in~\cref{fig:transition_paths} was obtained
by setting the origin of time at the LPT in well A for each trajectory,
and then averaging the coordinates as described above.

The most-likely trajectory shown in~\cref{fig:mb} was obtained by
computing the most-likely state $\alpha_j^\mathrm{ML}$ at each time $t_j$:
\begin{align}
  \alpha_j^\mathrm{ML} = \mathrm{argmax}_\alpha \left( \lbrace R_\alpha(t_j)
  \rbrace \right).
\end{align}

\section{Distribution of the transition-path time}
\label{sec:tpt_pdf}

We used theoretical results published in~\cite{Laleman2017}, investigating the
TPT for the symmetric one-dimensional quartic potential. For the
overdamped Langevin dynamics, the probability distribution function of the TPT reads:
\begin{widetext}
\begin{align}
  \begin{aligned}
    p_\mathrm{TPT}(t) &=  \sqrt{\frac{\beta E}{\pi}} \frac{2 \Omega}{1 - \mathrm{erf}(\sqrt{\beta
    E})} \ee^{-\Omega t} \left(1
    - \ee^{-\Omega t} \right)^{-\frac{3}{2}}\left(1
    + \ee^{-\Omega t} \right)^{-\frac{1}{2}} \exp{\left(-\beta E \frac{1 +
    \ee^{-\Omega t}}{1 - \ee^{-\Omega t}}\right)},
  \end{aligned}
  \label{eq:pexpr}
\end{align}
\end{widetext}
where $\beta E$ denotes the height of the potential barrier and $\Omega$
is proportional to the curvature at the top of the potential barrier. The previous
equation makes use of the error function:
\begin{align}
  \mathrm{erf}(x) = \frac{2}{\sqrt{\pi}} \int \limits_0^x \ud{u} \ee^{-u^2}.
\end{align}

The fits shown in~\cref{fig:tpt_fit} were performed by fitting $\beta E$
and $\Omega$ using the Levenberg-Marquardt algorithm.

\section{Transition rates for pancreatic cells}
\label{sec:pancreas}

We used a mouse pancreatic development
dataset~\cite{BastidasPonce2019}, and the RNA Velocity method
scVelo~\cite{Bergen2020} to obtain a transition matrix.
We normalized the raw counts data utilizing the default procedures from
scVelo. Each cell is size-normalized to the median of total molecules, and
\num{2000} top highly variable genes are chosen (with a minimum of \num{20} expressed counts for
spliced and unspliced mRNA). The scVelo pipeline calculates a
nearest neighbor graph using Euclidean distance utilizing \num{30} principal components on
logarithmic spliced counts, and the mean and variance for each cell its
calculated across is 30 nearest neighbors. We give below a Python
implementation using the scVelo package (\texttt{scv}), where
\texttt{adata} is the annotated pancreas dataset
from~\cite{BastidasPonce2019}.
\begin{verbatim}
scv.pp.filter_genes(adata, min_shared_counts=20)
scv.pp.normalize_per_cell(adata)
scv.pp.filter_genes_dispersion(adata, n_top_genes=2000)
scv.pp.log1p(adata)
scv.pp.moments(adata, n_pcs=30, n_neighbors=30)
\end{verbatim}

We then compute the velocity and transition matrix using scVelo-dynamical.
We used the following functions to recover the splicing kinetics of each
gene, and the cell-specific latent time which are estimated using
expectation-maximization. We computed the transition matrix based on the
velocity graph, using the scVelo function with default parameters.
\begin{verbatim}
scv.tl.recover_dynamics(adata)
scv.tl.velocity(adata, mode="dynamical")
scv.tl.velocity_graph(adata)
scv.utils.get_transition_matrix(adata, 'velocity')
\end{verbatim}


\clearpage
\renewcommand\thefigure{S\arabic{figure}}
\setcounter{figure}{0}

\begin{figure*}[h]
\centering
  \includegraphics[width = 3.5 in, page=6]{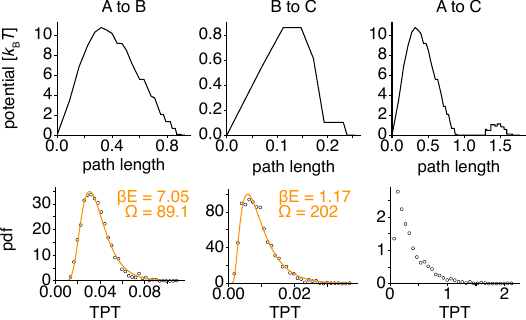}
  \caption{Supplementary figure to~\cref{fig:transition_paths} of the main text. Distribution of the TPT from A to B and from B to C for the M\"uller-Brown potential. The solid lines denote the fit to the theoretical prediction from~\cite{Laleman2017} (see~\cref{sec:tpt_pdf}).}
  \label{fig:tpt_fit}
\end{figure*}

\begin{figure*}[h]
\centering
  \includegraphics[width = \linewidth, page=7]{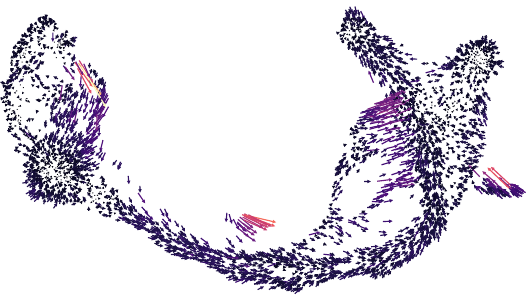}
  \caption{Supplementary figure to~\cref{fig:scrna} of the main text. Velocity field computed as $\vec{v}_\alpha = \sum_{\beta \ne \alpha} (\vec{x}_\beta - \vec{x}_\alpha) W_{\beta \alpha}$. Both the lengths and the colors of the arrows scale linearly with $\| \vec{v}_\alpha \|$.}
  \label{fig:scrna_velocities}
\end{figure*}

\end{document}